\documentclass{IEEEtran}
\pdfoutput=1
\usepackage{amsmath,amssymb,amsfonts}
\usepackage{algorithm}
\usepackage{algorithmicx}
\usepackage{algpseudocode}
\usepackage{graphicx}
\usepackage{textcomp}
\usepackage{subcaption}
\usepackage{multicol}
\usepackage{lipsum}
\usepackage{enumerate}
\usepackage{bm}
\usepackage{etoolbox}
\usepackage{array}
\usepackage[numbers]{natbib}
\newtheorem{lemma}{Lemma}
\newtheorem{theorem}{Theorem}
\newtheorem{corollary}{Corollary}
\newtheorem{definition}{Definition}

\newcounter{rowcntr}[table]
\renewcommand{\therowcntr}{\thetable.\arabic{rowcntr}}

\newcolumntype{N}{>{\refstepcounter{rowcntr}\therowcntr}c}

\AtBeginEnvironment{tabular}{\setcounter{rowcntr}{0}}

\newcommand{\RNum}[1]{\uppercase\expandafter{\romannumeral #1\relax}}

\def\BibTeX{{\rm B\kern-.05em{\sc i\kern-.025em b}\kern-.08em
    T\kern-.1667em\lower.7ex\hbox{E}\kern-.125emX}}
\begin{document}

\title{Intro-Stabilizing Byzantine Clock Synchronization in Heterogeneous IoT Networks
\thanks{This work has been submitted to the IEEE for possible publication. Copyright may be transferred without notice, after which this version may no longer be accessible.}
}

\author{Shaolin Yu*, Jihong Zhu, Jiali Yang, Wei Lu\\Tsinghua University, Beijing, China \\ysl8088@163.com}

\maketitle

\begin{abstract}
For reaching dependable high-precision clock synchronization (CS) upon IoT networks, the distributed CS paradigm adopted in ultra-high reliable systems and the master-slave CS paradigm adopted in high-performance but unreliable systems are integrated.
Meanwhile, traditional internal clock synchronization is also integrated with external time references to achieve efficient stabilization.
Low network connectivity, low complexity, high precision, and high reliability are all considered.
To tolerate permanent failures, the Byzantine CS is integrated with the common CS protocols.
To tolerate transient failures, the self-stabilizing Byzantine CS is also extended upon open-world IoT networks.
With these, the proposed intro-stabilizing Byzantine CS solution can establish and maintain synchronization with arbitrary initial states in the presence of permanent Byzantine faults.
With the formal analysis and numerical simulations, it is shown that the best of the CS solutions provided for the ultra-high reliable systems and the high-performance unreliable systems can be well integrated upon IoT networks to derive dependable high-precision CS even across the traditional closed safety-boundary.
\end{abstract}


\section{Introduction}
\label{sec:Introduction}
Clock synchronization (CS) is fundamental in designing traditional Distributed Real-Time Systems (DRTS) \citep{Kopetz1992Sparse,Kopetz1993TTP,Flexray,as6802,Dolev2014PulseGeneration} and today's Real-Time Embedded Systems (RTES) \citep{Kopetz2011Principles}, Cyber-Physical Systems (CPS), Wireless Sensor Networks (WSN), Internet of Things (IoT), and many other distributed systems.
In practice, by providing a \emph{sparse global time base} \citep{Kopetz1992Sparse,Kopetz2007WhySparse} for distributed applications \citep{Kopetz2011Principles}, not only the communication efficiency can be improved with statically optimized Time Division Multiple Access (TDMA) schedules \citep{Pozo2015schedules} but the system design and verification \citep{Steiner2011Formal,Sorea2008Verification} can be greatly simplified in comparison with that of asynchronous real-time systems \citep{Miller05nasa}.
Traditionally, as the required resources in providing the \emph{fault-tolerant} CS service are often expensive, only a small number of real-world DRTS systems (such as the high-end safety-critical systems in avionics) can acquire some kind of \emph{reliable} global time base.
While most of the other DRTS systems can only be built upon some unreliable CS schemes \citep{mills1991internet,ieee1588v2} or even be deployed under the traditional globally asynchronous architecture \citep{Chapiro84}.
With the rapid development of embedded computing and communication technologies, this situation changes drastically.
On one aspect, the required communication and computation in implementing the fault-tolerant CS become more and more affordable, with the progress of low-end embedded Commercial-Off-The-Shelf (COTS) devices such as the Ethernet, embedded processers, and Field Programmable Gate Array (FPGA).
On the other aspect, with the boost of everywhere computing and communicating, various modern DRTS systems are booming in accommodating ever-changing personal and social needs.
As a result, being built upon multi-scales networks comprised of Wide Area Network (WAN), Local Area Network (LAN), and Personal Area Network (PAN) \citep{Synchronization20205928IoT} with different networking technologies including traditional Ethernet, Software Defined Network (SDN), Time-Sensitive Networking (TSN), Software-Defined Radio (SDR), and even Radio Frequency Identification (RFID), these modern DRTS systems exhibit great diversity and complexity.
In this background, \emph{dependable} CS would play a more and more critical role in seamlessly integrating the trustworthy services for the diversified DRTS applications.

However, there is still a big gap between the dependability of the CS solutions provided in the emerging diversified DRTS and that provided in traditional DRTS.
At one extreme, high-end DRTS (like trains and civil aircraft) often requires the Mean-Time-To-Failure (MTTF) to be significantly better than $10^{9}$ hours \citep{Littlewood1993Dependability,Advances1994}.
To satisfy this, distributed CS systems are often built upon small-scale communication networks with statically connected homogenous components.
In this context, Byzantine-fault-tolerant \citep{PSL1980} CS  (BFT-CS) solutions \citep{Lamport1985Synchronizing,Welch1988New,OptimalToueg1987} are provided with the assumption that a fraction of the distributed components can fail arbitrarily \citep{Kopetz2003containment,Kopetz2004Hypothesis}, or saying, under the full control of a malicious adversary.
Further, self-stabilizing \citep{Dijkstra1974} BFT-CS (SS-BFT-CS) solutions \citep{DaliotBiological2003,DolevPulseBoundedDelay2007,Hoch2006SSBDCS,Ben2008FastSSBTDigital,LenzenOptimalCounting2015,Khanchandani2016Optimal,Dolev2016computational,Rybicki2016Near,Lenzen2019AlmostConsensus} are also provided with tolerating both transient system-wide failures and an amount of permanent Byzantine component failures.

At the other extreme, the CS schemes, such as Network Time Protocol (NTP) \citep{mills1991internet} and Precision Time Protocol (PTP) \citep{ieee1588v2}, referenced in the emerging IoT systems \citep{Synchronization20205928IoT,BOJIC2015361} often inevitably run in large-scale open environment (such as the Internet) with dynamically connected members.
In this context, the proposed CS solutions are seldom under the assumption of a non-cryptographic adversary or even just the common computationally limited attackers \citep{ullmann2009delay,Lisova2016ARP}.
Although some existing works deal with the attack-monitoring \citep{Lisova2016Protecting,Lisova2017synchronization} or reconfiguration problems \citep{Feldmann2020Overlay} upon open-world networks, the current results are far from being sufficient in concerning the possible far-reaching influence of future DRTS, especially the IoT systems \citep{Jha2020Threats}.
For example, with the ever-evolving communication technologies, today there are SDN, SDR, TSN, and various kinds of customized and non-standardized more intelligent switches and routers.
As more and more core network functions are developed with programmable and flexible devices such as embedded processors and Field-Programmable Gate Array (FPGA), the failure modes of these devices are more and more unpredictable.
However, malign faults are seldom considered in building practical IoT systems.
For another example, some industrial safety-critical applications can be attacked by open-world hackers and result in loss of control (as the recent accident encountered by the Colonial Pipeline \citep{Pipeline2021}).
Especially in considering that there might be a great number of safety-critical applications to be built upon various IoT systems, the internal operations of these systems should be safe enough.
In this respect, the dependability of CS solutions (such as the master-slave paradigm taken in PTP) proposed in the emerging IoT systems (might be with a massive number of sensors and actuators) is far below that of the SS-BFT-CS solutions taken in traditional DRTS.
And this would expose the IoT systems to risks of uncovered common malfunctions \citep{Estrela2012Challenges,Alghamdi2020Attacks}, undetected attacks, or even undesired emergences \citep{Kopetz2015emergence} in considering the so-called \emph{one-in-a-million} events \citep{Driscoll2003reality} or just the unknown intelligent invaders with limited computational resources.

\subsection{Motivation}
To mitigate this gap, we aim to provide CS solutions with both high reliability and high performance upon the emerging IoT networks.
Concretely, we would investigate the \emph{intro-stabilizing} (IS) BFT-CS problem upon IoT networks where some kind of external clocks are expected to be utilized while such kind of external clocks is not always reliable.
The so-called \emph{intro-stabilization} is extended from the traditional concept \emph{self-stabilization} \citep{Dijkstra1974} to provide discreet use of the external resources like the open-world reference clocks.
Meanwhile, the BFT-CS problem is investigated in sparsely connected low-degree IoT networks.
Also, by leveraging the existing CS schemes like the PTP as low-layer primitives, we expect that the advantages of the original CS schemes, such as hardware-optimized time precision and computational efficiency, can be inherited in the overall CS systems.
In presenting the IS-BFT-CS solution, we would also discuss the decoupled and easier error-detecting, correcting, fault-tolerant startup, and restartup procedures in the presence of the malicious adversary.
With this, we expect that the reliability, efficiency, and synchronization qualities of the CS systems can be better integrated by complementing traditional BFT-CS solutions with widely available time references given in the open world.

\subsection{Main obstacles}

In considering the overall problem, firstly, as real-world IoT networks are often across the WAN, LAN, and PAN areas \citep{Synchronization20205928IoT}, the first-of-all question is how a \emph{dependable} CS system can be deployed in such an all-scale network.
From the traditional viewpoints \citep{Kopetz2003containment,Kopetz2004Hypothesis,Driscoll2003reality}, current assumptions about the open-world adversaries might be overoptimistic.
For example, an unknown number of attackers arbitrarily distributed on the Internet may be very familiar with the provided CS algorithms.
Meanwhile, they can often well-disguise themselves to attack the target systems.
What is more, if these intelligent \emph{network neighbors} can attack some system from somewhere of the open-world network for a while, it is no reason to think that they would not attack it intermittently from elsewhere of the network.
In this situation, the attack-monitoring \citep{Dalmas2015ImprovingPTP,Lisova2016Protecting,Lisova2017synchronization} for the synchronization states and multi-source selection \citep{Estrela2014multisource} may be insufficient.
Notice that such \emph{worst cases} in the open world is much different from that in the closed world where all components are only exposed to \emph{physical} permanent failures and \emph{unintentional} system-wide transient failures within the strictly closed safety-boundary of the system.

Secondly, in considering malign faults in practical IoT systems, as we can hardly restrict the kinds of hardware devices, networking schemes, or low-layer protocols in developing the CS systems, the failure modes of the synchronization nodes can hardly be restricted.
In this context, it is safe to assume that these synchronization nodes can fail arbitrarily, for example, sending very different clock information and local states to different recipients.
Meanwhile, with the fault-independence assumption of distributed systems, it is unlikely that more than a fixed number of synchronization nodes are faulty at the same time in a real-world IoT system, providing that this system is operated in a distributed and closed way.
In this situation, it is often sufficient to assume that at most $f$ nodes are faulty arbitrarily in the $n$-node distributed synchronization system while all the other $n-f$ nodes are nonfaulty.
With this, the core problem is to provide the desired distributed services with the nonfaulty nodes in the presence of up to $f$ arbitrarily faulty nodes that are arbitrarily chosen and fully controlled by a malicious adversary.
This is in line with the core abstraction of the classical Byzantine General Problem (BGP \citep{Lamport1982Generals}).
In the literature (and also in this paper), the arbitrary faults that happened in the distributed nodes are referred to as the Byzantine faults.
Meanwhile, the distributed nodes being suffered from the Byzantine faults are referred to as the Byzantine nodes.

Thus, in the IoT networks, to validate the assumption that there are at most $f$ Byzantine nodes in the system, we do not allow the core BFT-CS algorithms to run across the WAN area.
Meanwhile, as the terminal devices (such as the sensors and actuators) deployed in the PAN area of the IoT systems \citep{Synchronization20205928IoT} are often energy-constraint (like the passive RFID tags), they can hardly be utilized as synchronization servers physically.
So, we only allow these low-power end devices to be passively synchronized, just like the thin clients and thick clients proposed in \cite{Sathiya2018Architecture}.
We see that there are several existing synchronization protocols such as Flooding Time Synchronization Protocol (FTSP) \citep{FTSP2004} and Timing-sync Protocol for Sensor Networks (TPSN) \citep{TPSN2003} aiming for synchronizing the low-power end devices with the \emph{edge nodes}, for example, the digital-twins \citep{Jia2021Twin}, of the upper-layer networks.
However, although these synchronization protocols pave the promising way for far-reaching observing, modeling, and controlling of the infinite physical world, they are mainly provided in the open-world wireless networks and take the master-slave paradigm, which cannot establish nor maintain the desired synchronization states of the system in the presence of the so-called Byzantine faults.
So, in viewing the big picture, it is urgent to build a reliable synchronization between the so-called \emph{edge nodes}.
Thus, we confine the main problem in this paper as to synchronize the devices in the LAN area with sufficient dependability, precision, and accuracy, while also providing a minimized safe interface in \emph{optionally} communicating with the upper-layer CS schemes and the lower-layer CS schemes.
With this minimized safe interface, the CS of LAN can be better integrated with the existing CS schemes of WAN and PAN.

Despite the whole problem, the confined LAN-layer CS problem is still nontrivial in IoT systems.
From the traditional viewpoints, one main obstacle in implementing a BFT-CS solution upon a practical LAN network is the insufficient connectivity of real-world communication infrastructures.
Namely, as is manifested in the classical Byzantine agreement (BA) problem \citep{Dolev1982StrikeAgain}, the network connectivity should be at least $2f+1$ in tolerating up-to $f$ Byzantine faults.
Alternatively, to mitigate this, practical high-end solutions \citep{Bauer2003guardian,as6802} also invest in designing specific hardware \emph{Byzantine filters} \citep{Driscoll2003reality}.
However, real-world LAN networks of IoT systems can hardly afford sufficiently high connectivity nor sufficiently designed \emph{Byzantine filters}.

Despite the limited network connectivity, there are also other obstacles.
Firstly, the required computation, storage, and communication in executing the BFT-CS algorithms often grow fast with the increase of the system scale.
As there can be a massive number of nodes being deployed in the IoT networks, high scalability of the BFT-CS solutions is desired.
Besides, as real-world communication infrastructures of IoT are diversified in physical interfaces (such as wired, wireless, optical) and technical standards (such as legacy Ethernet, Gbit Ethernet, SDN, TSN), an additional obstacle is that not all devices in the heterogeneous network can be directly connected.
Also, as the numbers of network interface controllers (NIC) in devices like the Ethernet switches are always bounded, only networks with bounded node-degrees can be provided.
Last but not least, the precision and accuracy required in the CS might be far below the maximal possible delay experienced in the IoT networks, which means that the basic fault-tolerant CS solutions provided in bounded-delay message-passing networks cannot be directly employed in IoT networks.

\subsection{New possibilities}
\begin{figure}[htbp]
\centerline{\includegraphics[width=2.7in]{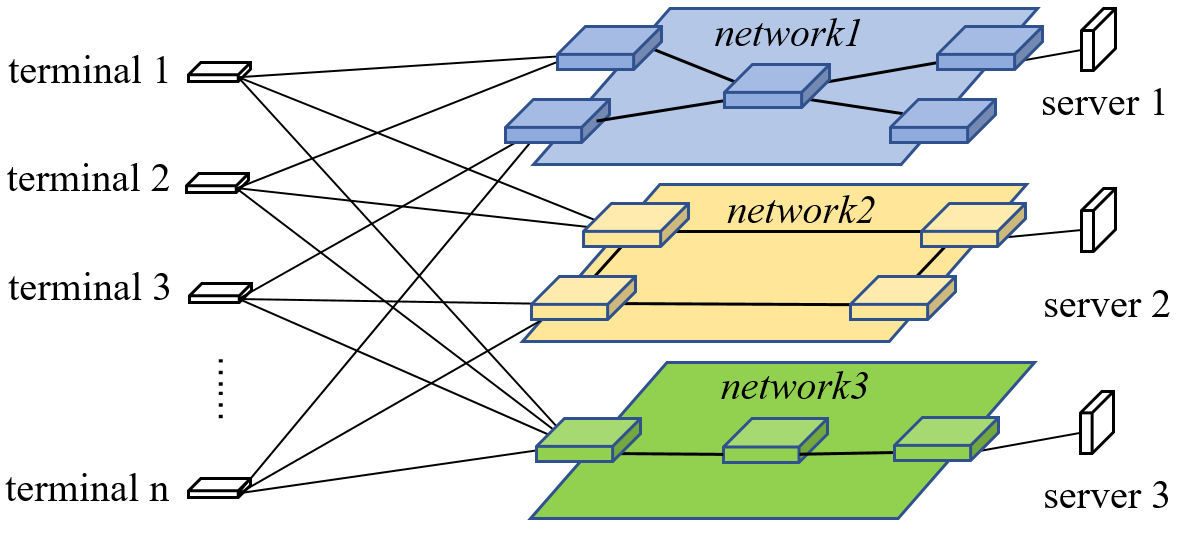}}
\caption{A typical heterogeneous LAN network.}
\label{fig:network}
\end{figure}

Nevertheless, there are also new possibilities.
Firstly, with today's modularized communication technologies, an embedded IoT device can be equipped with several NIC modules, such as the Wireless Fidelity (WIFI) module, the fast Ethernet module, and the Gbit Ethernet module, to perform diversified measuring, monitoring, and even modeling functions \citep{Jia2021Twin}.
In this background, these devices can often connect more than one kinds of communication infrastructures.
As is shown in Fig.~\ref{fig:network}, each \emph{computing device} (for example, the leftmost blocks) is allowed to communicate with more than one kind of \emph{bridge devices} (the colored blocks) in the typical real-world heterogeneous LAN network.
Following our former work \cite{TDWALDEN}, such \emph{computing devices} can be employed as multi-degree nodes in the LAN networks.

Secondly, unlike the traditional high-reliable CS solutions deployed in the \emph{fly-by-wire} \citep{Kopetz2004Hypothesis,Kopetz2011Principles,Steiner2008StartupRecovery} applications, the required overall weight, volume, and power supplies of the CS systems in IoT applications can be largely relaxed.
Also, the needed recovery time of the IoT systems can be largely relaxed in most real-world applications in comparison with that of avionics systems.
Moreover, as there are often various available external time references (such as the NTP and GPS clocks) in common IoT systems, various strategies can be proposed to utilize these external time references.
In this context, self-recovery is not required to be theoretically self-stabilizing but is expected to be more accessible, flexible, and still reliable.
For example, it is promising to seek ways to utilize available external time references while avoiding the intelligent attackers to leverage this as a new way to sabotage the system.
So, here the new problem is to efficiently synchronize the IoT networks with available external resources in the presence of various faults.

Besides, as is investigated in \cite{TDWALDEN}, some easy fault-tolerant operations can also be performed on the side of the \emph{bridge devices} (such as the customized Ethernet switches, SDN switches \citep{LaraNetwork}) or at least be performed on some embedded server node (the rightmost blocks in Fig.~\ref{fig:network}) being connected to each kind of communication infrastructure.
With this, each kind of communication infrastructure together with the server node connected to it can be viewed as an abstracted node (i.e., a single fault-containment region, FCR \citep{Kopetz2003containment,Kopetz2004Hypothesis,Steiner2013Interoperability}) in the LAN area.
By this trade-off, the original arbitrarily connected communication infrastructures can remain unchanged while the minimal network connectivity required in classical BFT-CS solutions can largely be supported to some extent in some kind of bounded-degree networks.

\subsection{Basic ideas and main contribution}
In this paper, we provide an IS-BFT-CS solution upon IoT networks where the communication infrastructures are heterogeneous, and the computing devices and the bridge devices are all sparsely connected (with bounded node-degrees).

Firstly, for the efficiency of networking, we only require that each kind of communication network be arbitrarily connected (which is also the minimum requirement in the original IoT networks), and there are more nonfaulty communication networks than faulty ones.
With this, as it is unlikely that there are more than a half number of the communication networks being faulty at the same time, the reliability of the overall CS system can be enhanced.
The basic idea is that, as we can deploy much more terminal nodes in the system than the available communication networks, the insufficient connectivity of the physical networks can be largely compensated by viewing the subnetworks as super nodes being inter-connected with a number of terminal nodes.
With this shrinking operation, the abstracted network would gain sufficient connectivity at the expense of increased failure rates of the super nodes.
Now by allowing almost half of the super nodes to fail arbitrarily, the networking problem and the fault-tolerance problem can be better balanced.

Secondly, for the high-quality and efficient CS, we employ the original CS schemes as synchronization primitives to achieve high synchronization precision without changing the underlying realizations of the primitives.
With this, the provided BFT CS algorithms can achieve synchronization precision in a similar order to the original CS schemes in the presence of Byzantine faults.
The basic idea is that, although synchronization precision provided in SS-BFT-CS solutions is often restricted by the maximal message delays, this precision can be further improved in stabilized CS systems by utilizing high-precision CS protocols like PTP as underlying primitives.
For this, as the stabilized CS system can provide well-separated semi-synchronous rounds, synchronous protocols such as the approximate agreement can be well simulated in a semi-synchronous manner with temporally well-separated remote clock readings.
So, with the basic convergence property of the approximate agreement, the synchronization precision can be in the same order as the bounded errors of the remote clock readings and bounded clock drifts.

Thirdly, for the efficiency of the IS-BFT-CS solution, the exact Byzantine agreement is avoided in establishing and maintaining the synchronization.
Moreover, the required stabilization time only depends on the number of the communication networks (denoted as $n_1$) and is independent of the number of the terminal devices (denoted as $n_0$).
Furthermore, once the system is stabilized, the complexity of computation, communication, and storage would be linear to $\max\{n_0,n_1\}$.
The basic idea is that, by constructing a closed safety boundary for the core CS system, the internal operations of the system within the closed safety boundary can be largely independent of the unknown open-world attacks.
With this, we can safely utilize some open-world time resources in the presence of possible attacks from open-world intelligent adversaries, as long as the adversaries cannot know when the open-world time resources are utilized.
Concretely, in the provided IS-BFT-CS solution, the open-world time resources are only utilized when the system is not stabilized.
This kind of property of the CS system is not much investigated in the existing works but may improve the reliability of real-world CS systems without adding great investments.

\subsection{Paper layout}
In the rest of the paper, the related work is presented in Section~\ref{sec:Related}, with emphasis on the integration of distributed BFT-CS provided for the ultra-reliable DRTS applications and the common master-slave CS provided for high-performance, high-precision, but unreliable applications.
The system abstraction of the considered IoT networks is given in Section~\ref{sec:Model}.
In Section~\ref{sec:NonSSBFTCS} and Section~\ref{sec:SSBFTCS}, the basic non-stabilizing BFT-CS and the basic IS-BFT-CS algorithms are successively introduced.
The worst-case analysis of these algorithms is presented in Section~\ref{sec:Analysis}.
In Section~\ref{sec:Result}, simulation results are also given in measuring the average performance of the IS-BFT-CS solution.
Finally, the paper is concluded in Section~\ref{sec:Conclusion}.

\section{Related works}
\label{sec:Related}
\subsection{Classical problem and solutions}
Dependable clock synchronization is a fundamental problem in building dependable DRTS applications.
Traditionally, as the certification authorities in the aviation industry demand convincible proof in showing the MTTF of the certified system being better than $10^9$ hours \citep{Kopetz2004Hypothesis,Advances1994,Littlewood1993Dependability}, significant efforts have been devoted to providing ultra-high reliable CS solutions.
To this end, as it is impossible to exhibit the desired system dependability by testing more than $100000$ years \citep{Littlewood1993Dependability}, distributed fault-tolerant methods are developed under the assumption that the MTTF of the independent hardware components might be with several orders of magnitude below (as can be experimentally observed) than that of the desired systems \citep{Kopetz2004Hypothesis}.
Under such assumptions, real-world distributed fault-tolerant systems are built by deploying sufficiently redundant subsystems \citep{Kopetz1993TTP,Miner2002ROBUS,Kopetz2003TTA,as6802}.
Moreover, as one cannot easily show the behaviors of the faulty subsystems being under some restricted patterns, it is often necessary \citep{Driscoll2003reality} to assume that the faulty subsystems can fail arbitrarily, i.e., being Byzantine \citep{Lamport1982Generals}.
In this context, classical BFT-CS algorithms are proposed in satisfying the dependability demanded in communities ranging from aviation, on-ground transportation, manufacturing industries, and other safety-critical realms \citep{Lamport1985Synchronizing,Dolev1986Impossibility,kopetz1987clock,Welch1988New}.

Besides the basic BFT, the CS algorithms running for the dependable DRTS applications are also required to be self-stabilizing \citep{Dijkstra1974} in tolerating transient system-wide failures caused by uncovered transient disturbances \citep{Kopetz2003containment} such as some severe interference like lighting \citep{Driscoll2003reality,Kopetz2011Principles} and other unforeseen environmental hazards.
Namely, after the arbitrary transient disturbance, as long as a sufficient number of DRTS components are not physically damaged, synchronization should still be globally established between the undamaged components within the desired stabilization time.
As all the variable values recorded in the RAM devices of the DRTS system can be arbitrarily altered during the transient disturbance, an SS-BFT-CS algorithm should work under all possible initial states of the system.
In this context, several deterministic SS-BFT-CS algorithms \citep{Daliot2006Linear,DolevPulseBoundedDelay2007,Lenzen2019AlmostConsensus} with linear stabilization time have been proposed upon completely connected networks (CCN).
Furthermore, to break the hard lower-bounds on the stabilization time and complexity of the message, probabilistic SS-BFT-CS solutions \citep{DolevWelchSelf2004,Dolev2011PulseGeneration,Dolev2014Rigorously,Dolev2014PulseGeneration,Khanchandani2016Optimal,Lenzen2019AlmostConsensus} are also explored.

\subsection{From theory to reality}
However, most real-world industrial SS-BFT-CS solutions \citep{Driscoll2003reality} are not built upon pure SS-BFT-CS algorithms.
For example, the Time-Triggered Architecture (TTA) \citep{Kopetz2003TTA} takes a light-weight SS-BFT startup procedure \citep{Steiner2006Startup,Steiner2008StartupRecovery,Saha2016Startup} where some kinds of hardware \emph{Byzantine filters} \citep{Driscoll2003reality}, such as the central guardians \citep{Bauer2003guardian,Steiner2008StartupRecovery} in the Time-Triggered Protocol (TTP) or monitor-pairs \citep{as6802} in Time-Triggered Ethernet (TTEthernet), are employed.
With this, the advantage is that the stabilization time and complexity of the CS algorithms can be reduced in accommodating the stringent requirement of avionics and automotive industries.
However, the expense is that the hardware \emph{Byzantine filters} should be implemented and verified very carefully in both the design and realization processes to show adequate assumption coverage.
Except for some high-end safety-critical applications, most common DRTS applications cannot afford such a delicate implementation.

Besides the SS-BFT startup problem, a more fundamental restriction in applying the classical BFT solutions in typical DRTS applications is the networking problem.
As most of the efficient SS-BFT-CS solutions \citep{DolevPulseBoundedDelay2007,Dolev2014PulseGeneration} are built upon CCN, real-world systems should provide sufficient network connectivity in simulating the original SS-BFT-CS solutions.
For this, the most straightforward networking scheme is to connect all the computing devices with a bus or a star topology \citep{Kopetz1993TTP,Flexray}.
Obviously, the disadvantage of such a naive solution is that the bus or the central bridge device in the star topology forms a single point of failure, which goes far from the original intention of distributed fault-tolerance.
A better networking scheme employs two stars or switches \citep{Bauer2001Assumptioncoverage,Steiner2008StartupRecovery,Steinbach2010Comparing} in eliminating the single point of failure.
However, such a basic redundancy can only tolerate benign failures of the bridge devices.
In the literature, there are also BFT solutions that tolerate Byzantine faults in both computing devices and bridge devices \citep{Yan1989Achieving,Wang1989Achieving}.
But these BFT solutions are often based upon special localized broadcast devices and synchronous communication networks and do not aim for solving the SS-BFT-CS problem.
In \cite{TDWALDEN}, an SS-BFT-CS solution that tolerates Byzantine faults in both computing devices and bridge devices is proposed with expected exponential stabilization time and relaxed synchronization precision.
So an interesting question is how to safely reduce the stabilization time with available external time resources in the open-world networks.

Lastly, in considering the synchronization precision, although classical BFT-CS solutions can provide some deterministic precision and accuracy under the assumption of bounded message delays and bounded clock drift rates, these original properties often need to be further optimized to support ultra-high synchronization requirements.
For example, some prototype solution \citep{Ademaj2007TTE1588} that integrates the time-triggered communication and the IEEE 1588 protocol \citep{ieee1588v2} exists in providing high synchronization precision for prototype TTEthernet, but without considering the BFT nor the self-stabilizing problem.
Later in the standard TTEthernet \citep{as6802}, such high synchronization precision is supported with hardware-supported transparent clocks \citep{as6802}.
However, restricted failure-mode of the Time-Triggered switches is required, which is then supposed to be supported with specially designed monitor-pairs (can be viewed as the hardware Byzantine filters \citep{Driscoll2003reality}).
Other high-precision CS solutions, such as the one provided in the White-Rabbit (WR) project \citep{Moreira2009White}, can even achieve sub-nanosecond precision by integrating both Synchronous Ethernet (SyncE) and PTP.
But it is only provided in the master-slave paradigm without considering malign faults.
In the extended PTP solutions \citep{Muhrextendingieee}, people also seek ways to enhance the reliability of PTP with redundant servers.
But these solutions are not for the Byzantine fault tolerance problem nor the stabilization (self-stabilization or intro-stabilization) problem.
As far as we know, there is no integration of SS-BFT-CS solution and IEEE 1588 upon sparsely connected network in DRTS applications without assuming some components generating benign faults only.

\subsection{The missing world for synchronizing IoT}
We can see that, for the CS problem, although the communication infrastructures of IoT are not better than that of traditional DRTS, they are not much worse, especially in the LAN area.
But existing CS schemes proposed for IoT (such as PTP) are mainly derived from the server-client paradigm (including the master-slave one, the same below) proposed for the Internet and WSN, while seldom from the distributed paradigm proposed for traditional DRTS.
However, the server-client CS schemes adopted on the Internet, such as the NTP \citep{mills1991internet} and Simple NTP (SNTP) \citep{SNTP2006}, are not intentionally provided for real-time applications and can only provide best-effort services with coarse time precision.
Meanwhile, the CS schemes provided for the WSN, such as the FTSP \citep{FTSP2004}, TPSN \citep{TPSN2003} and other wireless synchronization protocols \citep{Sathiya2018Architecture,Jia2020Clustering,Zhou2020Carrier,Jia2021Twin}, are mainly for large-scale dynamical networks consisting of tiny wireless devices with strictly restricted power-supply and physical communication radius.
Besides, these CS schemes are provided mainly for real-time measurements but not for hard-real-time controls like the CPS applications.
As a result, most of these CS schemes cannot tolerate Byzantine faults of some critical servers, masters, or other kinds of \emph{central} nodes.
This would gravely restrict the reliability of the emerging far-reaching large-scale IoT systems.
For a simple example, some middle-layer NTP servers deployed in the CS systems may be attacked by some stealthy attackers (hard to detect) to send and relay inconsistent messages to all other nodes.
However, the receivers cannot always distinguish the faulty messages from the correct ones without employing Byzantine fault-tolerance.
Viewing the CS solutions provided for the Internet and the WSN as vivid instances of social world synchronization and physical world synchronization, respectively, we see a missing link between these two ultimate worlds in looking forward to the future dependable IoT applications.
But unfortunately, this cannot be fixed by only adopting some other kind of server-client solutions, such as gPTP \citep{ieee8021as} and ReversePTP \citep{Mizrahi2016REVERSEPTP}.

To mend this, just between the social world where the members are intellectually unrestricted and the physical world where the devices are physically restricted, there might be a better place where certainties can be built upon firm realistic foundations.
Namely, in the words of the multi-layer networks, the internal CS (ICS) in the LAN should be as dependable as possible to minimize the influence of uncertainties raised from both the WAN and PAN sides.
In this context, the main problem is to provide efficient high-reliable ICS upon the LAN networks of IoT while maintaining the advantages (high-precision, low-complexity, low-cost, etc.) of the original unreliable CS protocols (such as PTP or even the ultra-high-precision WR).
Also, as external time is often available in the IoT systems, some kinds of external time references may be helpful.
Further, providing that the ICS systems can be well designed, the remaining problem is integrating these systems with external CS (ECS).
For this, integrations of ICS and ECS are provided in the literature \citep{Cristian1995external,Fetzer1997Integrating,Kopetz2004Integration}.
But up to now, with our limited knowledge, the SS-BFT (and IS-BFT) ICS solution upon heterogeneous IoT networks is still missing.

\section{System model and the main problem}
\label{sec:Model}
In this section, we give a basic model to characterize the discussed heterogeneous IoT network in handling the related CS problem.
Generally, the whole IoT system $\mathcal{N}$ is constituted by three kinds of subsystems: the WAN systems, the LAN systems, and the PAN systems.
For the confined CS problem, we first introduce the LAN system and then briefly introduce its interfaces to the other two kinds of systems.

\subsection{The LAN system}
\label{subsec:LANModel}
As is presented in Fig.~\ref{fig:network}, an LAN system (denoted as $\mathcal{L}$) consists of $n_0\geqslant 6$ terminal nodes (denoted as $i\in V_0$ with $V_0=\{1,\dots,n_0\}$) and a heterogeneous bridge network $\mathbf{G}$.
The heterogeneous bridge network $\mathbf{G}$ is comprised of $n_1\geqslant 3$ disjoint (homogeneous) bridge subnetworks, denoted as $G_s\in \mathbf{G}$ for $s\in S=\{1,\dots, n_1\}$.
Each such bridge subnetwork $G_s=(B_s,E_s)$ consists of $|B_s|$ connected bridge nodes, each denoted as $b_{s,q}\in B_s$, and $|E_s|$ bidirectional communication channels.
As $\mathbf{G}$ is heterogeneous, the bridge nodes $b_{s_1,q_1}$ and $b_{s_2,q_2}$ cannot be directly connected whenever $s_1\neq s_2$.
The terminal nodes can be connected to the bridge nodes with bidirectional connections (denoted as $E_0$) but with the node-degree of every terminal node being no more than $d_0$.
Also, the node-degree of every bridge node is no more than $d_1$.
Thus, the network topology of $\mathcal{L}$ is a bounded-degree undirected graph, denoted as $\mathbf{H}=(V_0\cup B_1 \cup \dots \cup B_{n_1},E_0 \cup E_1 \cup \dots \cup E_{n_1})$.
Generally, the bridge network $\mathbf{G}$ can also be wholly or partially homogeneous.
Here we consider the worst cases.
Practically, as the number of the communication infrastructures is often limited, we assume $n_1$ is a fixed number equal to or greater than $3$.
For simplicity, we assume $d_0=n_1$ and each $i\in V_0$ is a synchronization server node being directly connected to the $n_1$ bridge subnetworks.
It is obvious that $\mathbf{H}$ can be extended with an $O(\log n_0)$ diameter for any $d_1\geqslant 3$.

In providing backward compatibility, we assume that each bridge subnetwork $G_s$ is directly connected to a network-manager node $s\in S$ with a bidirectional communication channel (as the server nodes in Fig.~\ref{fig:network}).
The terminal nodes $V_0$ and the network-manager (\emph{manager} for short) nodes $S$ are all referred to as the computing nodes, as they can perform the required computation.
The bridge nodes in a nonfaulty $G_s$ can deliver the messages between the manager node $s$ and the terminal nodes directly connected to $G_s$ following the underlying CS protocol $\mathcal{P}$ and communication protocol $\mathcal{C}$.
When considering babbling-idiot failures \citep{Kopetz2003containment} of the terminal nodes, the bridge nodes are assumed to be able to perform some rate-constrained communication for the incoming messages from the terminal nodes.
Concretely, $\mathcal{P}$ and $\mathcal{C}$ can be respectively interpreted as PTP (or even WR) and some rate-constrained Ethernet (such as IEEE AVB \citep{ieeeAVB}, AFDX \citep{Schneele2012AFDX}, TTEthernet \citep{as6802}, OpenFlow \citep{LaraNetwork}, TSN \citep{ieee8021as}) or other customized protocols.

In considering BFT of the terminal nodes, we assume up-to $f_0$ nodes in $V_0$ can fail arbitrarily since the real-time instant $t=t_{0}$.
For simplicity, the real-time $t$ is assumed to be a universal physical time, such as the Newtonian time.
And if not specified, the discussed \emph{time}, \emph{instants}, \emph{durations} and \emph{time intervals} are all referred to the real-time.
For our purpose, we assume the system is in an arbitrary state at $t_{0}$, and we only discuss the system since $t_{0}$.
With this, if a terminal node is not a Byzantine node, it is a nonfaulty node that always behaves according to $\mathcal{P}$, $\mathcal{C}$, and the provided upper-layer CS algorithms.
Besides, as the failures of the communication channels between the computing nodes and the bridge nodes can be equivalent to the failures of the computing nodes, the communication channels between them are assumed reliable.

In considering BFT of the bridge nodes and the manager nodes, as we allow that the bridges in each bridge subnetwork can be arbitrarily connected, each bridge subnetwork $G_s$ together with the manager node $s$ are deemed as a single FCR.
Concretely, a bridge subnetwork $G_s$ is nonfaulty during a time interval $[t,t']$ if and only if (iff) all bridge nodes and the internal communication channels in $G_s$ are nonfaulty during $[t,t']$.
We say a bridge node $b$ being nonfaulty during $[t,t']$ iff $b$ correctly delivers the messages during $[t,t']$.
In supporting the bounded-delay model \citep{DolevPulseBoundedDelay2007}, to correctly deliver a message $m$ in $b$, $b$ is required to deliver $m$ within a bounded message delay $\delta_\mathtt{q}$ in executing $\mathcal{P}$ and $\mathcal{C}$.
Practically, this bounded-delay requirement can be easily supported with rate-constrained Ethernet or even traditional Ethernet under low traffic loads \citep{Boudec2000calculus,Loeser2004LowLatencyEthernet,Loeser2004HardRTEthernet,Ademaj2007TTE1588}.

For CS, firstly, we assume that each nonfaulty computing node $i$ is equipped with a hardware clock $H_i$.
To approximately measure the time, each $H_i$ can generate ticking events with a nominal frequency $1/T_H$, where $T_H$ is the nominal ticking cycle.
As the accuracy of real-world clocks is imperfect, the actual ticking cycles of $H_i$ are allowed to arbitrarily fluctuate within the range $[(1-\rho)T_H,(1+\rho)T_H]$, where $\rho\geqslant 0$ is the maximal drift-rate of the hardware clocks.
At every instant $t$, the nonfaulty node $i$ can read the hardware clock $H_i$ as the number of the counted ticking events, denoted as $H_i(t)$ and referred to as the hardware-time of $i$ at $t$.
In considering the stabilization problem, $H_i(t_0)$ is assumed to take arbitrary values in a finite set $[[\tau_{max}]]$, where $[[x]]=\{0,1,\dots,x-1\}$ is the set of the first $x$ nonnegative integers.
And since $t_0$, $H_i(t)$ would not be written outside the hardware clock and would monotonically increase with respect to $t$ in counting the ticking events when $H_i(t)<\tau_{max}-1$.
When $H_i(t)=\tau_{max}-1$, $H_i$ would return to $0$ in counting the next ticking event and then continue to count the following ticking events.
As $H_i$ is read-only, it can be used for realizing the timers with fixed timeouts.
In performing clock adjustments in executing the CS algorithms, other kinds of clocks should be defined.
For simplicity, the value of the local clock $C_i$ at instant $t$ can be defined as $C_i(t)=(H_i(t)+\textit{offset}^{C}_i(t))\bmod \tau_{max}$, where $\textit{offset}^{C}_i(t)$ is the value of the local-offset variable $\textit{offset}^{C}_i$ at $t$.
In executing the CS algorithms, the local-time $C_i(t)$ is allowed to be read (or saying $C_i$ being used as input) at any $t$ by the $\mathcal{P}$ protocol running in $i$.
Also, $C_i(t)$ is allowed to be written (or saying $C_i$ being adjusted) at any $t$ by the CS algorithms running in $i$.
With this, the basic accuracy of $H_i(t)$ can be shared in $C_i(t)$ while the timers and the adjustments of the local clocks are decoupled.

Sometimes, we also need one or more kinds of logical clocks for convenience.
For example, by defining the logical clock of node $i$ as $L_i(t)=(C_i(t)+\textit{offset}_i(t))\bmod \tau_{max}$, $L_i(t)$ is called the logical-time of $i$ at $t$.
Here, the difference of the logical-time and the local-time of $i$ is represented as the logical-offset variable $\textit{offset}_i$ in $i$.
In this way, the basic accuracy and synchronization precision of $C_i(t)$ can be shared in $L_i(t)$ while the unnecessary coupling between the $\mathcal{P}$ and the upper-layer CS algorithms can be avoided.
It should be noted that the upper-layer CS algorithms are not completely decoupled with the underlying $\mathcal{P}$ protocol, as we allow the upper-layer CS algorithms to adjust $C_i$ instead of $L_i$ (or equivalently, we allow the underlying $\mathcal{P}$ protocol to use $L_i$ instead of $C_i$ as its input).
But such coupling is made as small as possible and can be supported in real-world realizations such as the common embedded systems.
Besides the $L$ clocks, other kinds of clocks can also be defined upon the local-time $C_i(t)$ or directly upon the hardware-time $H_i(t)$.
For example, we can define some \emph{alien} clock of node $i$ as $Y_i(t)=(H_i(t)+\textit{offset}^{Y}_i(t))\bmod \tau_{max}$ (can be specifically called the alien-time).
In considering the stabilization problem, all the offset variables for the clocks can be arbitrary valued in $[[\tau_{max}]]$ at $t_0$.
For convenience, as the hardware-times, local-times, logical-times, and alien-times are all circularly valued in $[[\tau_{max}]]$, we define $\tau_1\oplus \tau_2=(\tau_1+ \tau_2)\bmod \tau_{max}$ and $\tau_1\ominus \tau_2=(\tau_1-\tau_2)\bmod \tau_{max}$.
And to measure the difference of two such times $\tau_1$ and $\tau_2$, we define $\mathring{d}(\tau_1,\tau_2)=\min\{\tau_1\ominus \tau_2,\tau_2\ominus \tau_1\}$.

On the whole, by viewing each bridge subnetwork $G_s$ together with the corresponding manager node $s$ as an abstracted bridge node $j\in V_1$ ($V_1=\{n_0+1,\dots,n_0+n_1\}$), $\mathbf{H}$ can be further simplified as a completely connected bipartite network (CCBN) $G=(V, E)$ with $V=V_0\cup V_1$ and $E$ making the complete bipartite topology $K_{n_0,n_1}$.
An abstracted bridge node $j\in V_1$ is nonfaulty iff $G_s$, $s$, and the communication channels between them are nonfaulty.
The failures of the edges in $E$ are equivalent to the failures of the nodes in $V_0$.
With this, we assume that up-to $f_0=\lfloor (n_0-1)/5\rfloor$ terminal nodes in $V_0$ and $f_1=\lfloor (n_1-1)/2\rfloor$ abstracted bridge nodes in $V_1$ can fail arbitrarily since $t_{0}$.
All faulty nodes in $V_0$ and $V_1$ are denoted as $F_0$ and $F_1$, respectively.
The nonfaulty nodes are correspondingly denoted as $U_0=V_0\setminus F_0$, $U_1=V_1\setminus F_1$ and $U=U_0\cup U_1$.
As the network diameter of each $G_s$ can be bounded within $O(\log n_0)$, the overall delay of a message from a node $i\in U_0$ to a node $j\in U_1$ (and vice versa) can be bounded within $2\delta_\mathtt{p}+O(\log n_0) \delta_\mathtt{q}$, where $\delta_\mathtt{p}$ is an upper-bound of the processing delay for every message in every nonfaulty computing node.
For convenience, we assume the maximal overall message delay between $i$ and $j$ is less than $\delta_\mathtt{d}$.
For discussing CS upon the abstracted CCBN $G$, the clocks of each $s\in S$ are also used as the clocks of the corresponding node $j\in V_1$.
For convenience, we use $\mathtt{s}(j)=j-n_0$ to denote the corresponding manager node that is abstracted in $j$.
Also, for every $s\in S$, we use $\mathtt{s}^{-1}(s)=s+n_0$ to denote the corresponding abstract node $j\in V_1$.
This is only for strictly differentiating $j$ and $s$ in avoiding possible confusion.
No algorithm really needs to compute $\mathtt{s}(j)$ nor $\mathtt{s}^{-1}(s)$.
Similarly, we also define $\mathtt{s}(V')=\{\mathtt{s}(j)\mid j\in V'\}$ and $\mathtt{s}^{-1}(S')=\{\mathtt{s}^{-1}(s)\mid s\in S'\}$ for every $V'\subseteq V_1$ and $S'\subseteq S$, respectively.

Upon existing works \citep{Boudec2000calculus,Loeser2004LowLatencyEthernet,Loeser2004HardRTEthernet,Ademaj2007TTE1588,ieeeAVB,as6802,steinbach2012tomorrow,Schneele2012AFDX,LaraNetwork,ieee8021as,Steiner2016Next}, the given assumptions can be practically supported with today's COTS devices commonly used in IoT networks.
Also, it is often easier to add more terminal nodes than to add more communication networks in the IoT networks.
By allowing $n_0>5f_0$ and $n_1>2f_1$, the minimal realization of the IS-BFT-CS system only requires $n_1=3$, which is easier to be supported in real-world systems than the minimal requirement of deterministic BA (DBA) upon CCN.

\subsection{The interfaces for the two sides}

\begin{figure}[htbp]
\centerline{\includegraphics[width=3.0in]{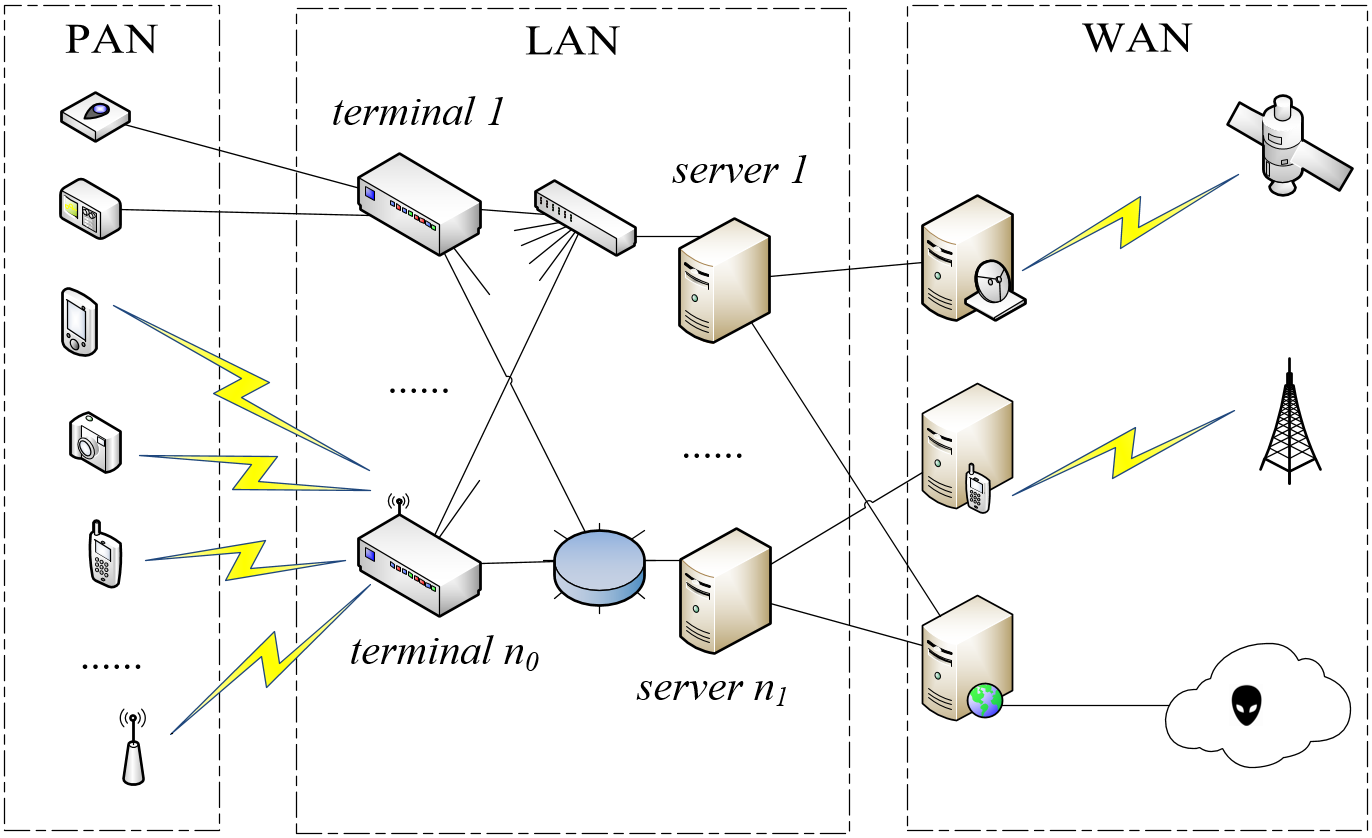}}
\caption{The external interfaces of the LAN network.}
\label{fig:network2}
\end{figure}

In the IoT system $\mathcal{N}$, the LAN system $\mathcal{L}$ should connect to one or more lower-layer PAN systems for interconnecting the \emph{things}.
Moreover, $\mathcal{L}$ is often connected to one or more higher-layer WAN networks for interconnecting of \emph{more things}, as is shown in Fig.~\ref{fig:network2}.

For the lower-layer side of $\mathcal{L}$, each terminal node $i\in V_0$ in the network $\mathbf{H}$ can serve as a synchronization server for the connected PAN nodes which serve as synchronization clients.
These PAN nodes can be low-power receivers, mobile stations, or even in-hand or wearable devices with dynamic accesses.
Each terminal node $i\in V_0$ can connect to more than one PAN network for scalability.
In the overall synchronization system, the communication between the terminal nodes in $V_0$ and the PAN nodes is unidirectional.
Namely, each nonfaulty terminal node $i\in U_0$ periodically broadcasts its current clock to the connected PAN nodes.
Meanwhile, the messages from the PAN nodes are all ignored by $U_0$ in the synchronization system.

For the upper-layer side of $\mathcal{L}$, firstly, each manager node $s\in S$ in the network $\mathbf{H}$ can be configured as a synchronization client for the connected WAN nodes.
These WAN nodes, denoted as $Z$ with $|Z|=n_2$, serve as \emph{time-abundant} external synchronization stations.
Namely, each node $z\in Z$ can access at least one kind of external time (UTC, TAI, etc.) with well-configured timing devices (such as GPS receivers, PTP clients, or just NTP clients), providing that the node $z$ is nonfaulty.
For simplicity and without loss of generality, we assume the external time is represented as the universal physical time $t$.
And $z\in Z$ is nonfaulty during $[t_1,t_2]$ iff every connected nonfaulty manager node $s\in S$ always reads the reference clock of $z$ (denoted as $R_z(t)$) with $\forall t\in [t_1,t_2]:\hat{R}_{z,s}(t)\in [t-e_0,t+e_0]$, where $e_0$ is the external time precision.
In the overall CS system, each $z\in Z$ can connect to more than one LAN network (like $\mathcal{L}$) for scalability.

Now at the side of $\mathcal{L}$, each $s\in S$ is typically connected to one node in $Z$.
Each $s$ can also connect to more than one node in $Z$ to tolerate some permanent faults that happened in $Z$ (such as shown in Fig.~\ref{fig:network2}).
Obviously, if more than one-half of the nodes in $Z$ is always nonfaulty, the BFT-CS problem is trivial by taking the majority from the timing information given by $Z$ in every nonfaulty $s\in S$.
In this case, we also say that the external time is available in $s$.
However, as this timing information is from the open world, we cannot ensure that a sufficiently large number of nodes in $Z$ would always withstand all \emph{intelligent} attacks from the open world.
So the external time is not always available in $s$.
This differs from the transient failures that should be tolerated with self-stabilization in traditional DRTS.
Namely, with the more realistic consideration of the open-world malignity, the intelligent attacks might be launched with an arbitrary frequency and deliberately designed intermittent periods.
Here, to differentiate it from the traditional \emph{self-stabilization} problem and the \emph{Byzantine General} problem, we can view the open-world time references in the overall synchronization problem as some resources in some \emph{Dark Forest}\citep{Cixin2016DarkForest}.
Namely, the so-called \emph{Dark Forest}\citep{Cixin2016DarkForest} might be a good (but sometimes being regarded as over-permissive) metaphor of the open-world resources (the forest) along with the unknown dangers (the darkness).
We argue that this kind of problem is not well handled in the open world and it might also be over-optimistically neglected in the emerging large-scale IoT systems.

In the context of the \emph{Dark Forest}\citep{Cixin2016DarkForest}, the nodes in $S$ should not always depend on the open-world timing information to update their clocks.
Instead, at every instant $t$, each nonfaulty $s\in S$ should select a subset $Z_s(t)\subseteq Z$ to decide its current time servers.
And when $Z_s(t)=\emptyset$, it indicates that $s$ does not use any timing information given by $Z$ at $t$.
So, a pure ICS solution is provided if $Z_s(t)=\emptyset$ always holds for every nonfaulty $s\in S$ and every $t$, just as the traditional ICS solutions.
And an external-time-based ICS solution is provided if $Z_s(t)=\emptyset$ holds whenever the system is stabilized while $Z_s(t)$ can be nonempty when the system is not stabilized.
In considering the dependability of the CS system in the context of the \emph{Dark Forest}, the provided IS-BFT-CS solution is an external-time-based ICS solution.
In this vein, the clocks $Y_{\mathtt{s}^{-1}(s)}$ derived from $\hat{R}_{z,s}(t)$ for all $s\in S$ and $z\in Z$ are called the \emph{alien} clocks.
When the external time is available in $s$, we also say $Y_{\mathtt{s}^{-1}(s)}$ is available.

\subsection{The underlying protocols}
To the underlying CS protocol $\mathcal{P}$, we assume that if two nonfaulty nodes $i$ and $j$ are connected by a nonfaulty bridge subnetwork $G_s$, $j$ can synchronize $i$ with $\mathcal{P}$ upon $G_s$ and vice versa.
Concretely, suppose that a point-to-point CS instance of $\mathcal{P}$, denoted as $\mathcal{P}_{j,i}$, runs between a server node $j\in U$ and a client node $i\in U$ since $t_{0}$ and no other instance of $\mathcal{P}$ runs between $i$ and $j$ nor any adjustment of $C_j$ happens.
Then, by running $\mathcal{P}_{j,i}$ in the server node $j$ and the client node $i$, $i$ can remotely read the local clock $C_j(t)$ as $\hat{C}_{j,i}(t)$.
And if for all $t\in [t_{0}+\Delta_{0},+\infty)$
\begin{eqnarray}
\label{eq:p_precision}\mathring{d}(\hat{C}_{j,i}(t)-C_j(t))\leqslant \varepsilon_{0}
\end{eqnarray}
holds, we say $\mathcal{P}$ is with the synchronization precision $\varepsilon_0$ and a stabilization time $\Delta_{0}$ (which includes the time for establishing the master/slave hierarchy and establishing the master-slave synchronization precision).
Further, if for all $\delta\leqslant\Delta$
\begin{eqnarray}
\label{eq:p_accuracy}|(\hat{C}_{j,i}(t+\delta)\ominus\hat{C}_{j,i}(t))-\delta|\leqslant\varrho_0\delta+\varepsilon_{0}
\end{eqnarray}
also holds, we say $\mathcal{P}$ is with the accuracy $\varrho_0$ for $\varepsilon_0$ and $\Delta$.
For $\mathcal{P}_{j,i}$, we assume $\varepsilon_0$ and $\Delta_{0}$ are all fixed numbers specified by the concrete realization of $\mathcal{P}$.
And as no adjustment of $C_j$ happens, the accuracy $\varrho_0$ of $\mathcal{P}$ can be no worse than $\rho$ for $\varepsilon_0$ and some $\Delta\approx\tau_{max}$ (slightly less than $\tau_{max}$, the same below).
In considering Byzantine faults, if $j$ is faulty, $\hat{C}_{j,i}(t)$ would be an arbitrary value in $[[\tau_{max}]]$ at any given $t$.
Here, the nodes $i$ and $j$ can be arbitrary computing nodes that are directly connected to a bridge subnetwork.

\begin{figure}[htbp]
\centerline{\includegraphics[width=1.6in]{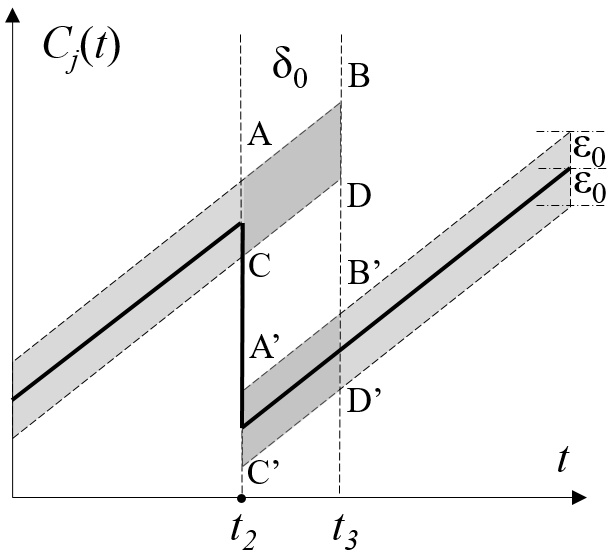}}
\caption{An updating span of $C_j$.}
\label{fig:adjustments}
\end{figure}

In considering adjustments of $C_j$, for simplicity, we assume that the $\mathcal{P}$ protocol updates the remote clocks with the instantaneous adjustments rather than the continuous adjustments.
Namely, when $j\in U$ and an adjustment of $C_j(t)$ (shown as the solid curve in Fig.~\ref{fig:adjustments}) happens at $t_{2}$, although there can be a period $[t_{2},t_{3}]$ during which $C_j(t)$ might be measured in node $i\in U$ as a value $\hat{C}_{j,i}(t)$ being arbitrarily distributed in the intersection of a vertical line and the two disjoint grey regions $ABCD$ and $A'B'C'D'$, this value cannot be inside the white region $CDA'B'$ at any given $t\in [t_{2},t_{3}]$.
Calling $[t_{2},t_{3}]$ as an updating span of $C_j$, for every such updating span, we require that the updating duration $t_{3}-t_{2}$ is bounded by $\delta_{0}$, after which (\ref{eq:p_precision}) and (\ref{eq:p_accuracy}) should hold until the beginning of the next updating span.
This requirement can be supported in most real-world hardware PTP realizations.
Also, we note that the realizations of $\mathcal{P}$ with continuous adjustments can also be accepted in the $\mathcal{P}$-based CS algorithms provided in this paper.
But to our aim, as the clock-updating time-bound $\delta_{0}$ should be as small as possible for reaching faster stabilization, instantaneous updating is preferred as it can often be much faster than the continuous one.
Also, as we should consider the worst-case performance of the CS solution, software optimization of the synchronization precision would not much help.

Lastly, to the underlying communication protocol $\mathcal{C}$, for every $i,j\in U$, $\mathtt{s}(j)$ can correctly communicate with $i$ by sending messages to $G_s$ and vice versa.
In the abstracted CCBN, every node $j\in U_1$ can send arbitrary message $m$ to every $i\in V_0$ at any instant $t\geqslant t_{0}$.
For efficiency, $j$ can also broadcast $m$ to all nodes in $V_0$.
When $i\in U_0$ receives such a message $m$, $i$ can deduce the sender of $m$ in $V_1$ with the connected communication channels.
Also, every $j\in U_1$ can deduce the sender of $m$ in $V_0$ with the nonfaulty bridge network $G_{\mathtt{s}(j)}$ and the fixed communication ports.
The messages can be signature-free, just like the unauthenticated messages sent in standard Ethernet, but should be with bounded frequencies and bounded lengths.

\subsection{The synchronization problem}
Now assume $n_0>5f_0$, $n_1>2f_1$, and there are no more than $f_0$ and $f_1$ Byzantine nodes in respectively $V_0$ and $V_1$ since $t_{0}$ (at which the system $\mathcal{L}$ can be with arbitrary initial system state).
Then, the nodes in $U$ should be synchronized with the desired synchronization precision $\varepsilon_{1}$ and accuracy $\varrho_{1}$ upon $G$ since $t_{1}$, where the actual stabilization time $t_{1}-t_{0}$ is expected to be sufficiently small.
Concretely, for the distributed CS, we say the $X$ clocks ($X$ can be $C$, $L$, or $Y$) of $P$ are $(\varepsilon,\varrho,\Delta)$-synchronized during $[t_{1},t_{2}]$ iff
\begin{eqnarray}
\label{eq:s_precision_general}\mathring{d}(X_i(t)-X_j(t))\leqslant \varepsilon\\
\label{eq:s_accuracy_general}|(X_i(t')\ominus X_i(t))-(t'-t)|\leqslant \varrho_{1} (t'-t) +\varepsilon
\end{eqnarray}
hold for all $i,j\in P$ and all $t',t\in [t_{1},t_{2}]$ with $0\leqslant t'-t\leqslant\Delta$.
With this, it is required that the $C$ clocks (and thus the $L$ clocks) of $U$ should be $(\varepsilon_{1},\varrho_{1},\Delta)$-synchronized during $[t_{1},+\infty)$ with some $\Delta\approx\tau_{max}$.
And when this happens, we say $\mathcal{L}$ is $(\varepsilon_{1},\varrho_{1})$-synchronized (and also stabilized) with the stabilization time $\Delta_{1}=t_{1}-t_{0}$.
As the $X$ clocks used in this paper are all with the same value range $[[\tau_{max}]]$, $\Delta\approx\tau_{max}$ can be a common parameter in all cases.
So for simplicity we say the $X$ clocks are $(\varepsilon,\varrho)$-synchronized when the $X$ clocks of $U$ are $(\varepsilon,\varrho,\Delta)$-synchronized.
To avoid DBA, we do not always require the stabilization time being a deterministically fixed duration.
Instead, a randomized stabilization time with an acceptable expectation $\Delta_{1}$ is also allowed.

In the context of the IoT networks, as the alien clocks are \emph{often} but not \emph{always} available, we should seek some discreet ways to integrate the ICS system with the alien clocks.
By assuming that the failures of the nodes in $\mathcal{L}$ are independent of those of the alien clocks, the new problem posed here is to construct some more efficient complementary system to integrate the closed-world resources with the open-world resources.
The real-world scenario is that, with the minimized safe interface of $\mathcal{L}$, the failures that happened in the ICS system can be largely assumed to be independent of that of the alien clocks.
Meanwhile, as the external time sources are often maintained in good condition, and the external attacks can often be promptly detected and handled with attack-monitoring \citep{Lisova2016Protecting,Lisova2017synchronization}, the alien clocks can be available most of the time.
So, when the ICS system experiences some transient system-wide failures (often caused by improper internal operations or some temporary device malfunctions), the probabilities of unavailable alien clocks are low.
Thus, this kind of availability of the alien clocks can be leveraged to integrate traditional ICS and the open-world time resources more discreetly.

\section{Non-stabilizing BFT-CS algorithms upon $G$}
\label{sec:NonSSBFTCS}
In this section, we first provide some non-stabilizing BFT-CS algorithms built upon some particular initial system states.
Then we will use some of these algorithms as building blocks for constructing the IS-BFT-CS solution in the following section.
For simplicity, we will prefer the abstracted nodes $V_1$ to the manager nodes $S$ in describing the algorithms running in the abstracted bridge nodes, although the algorithms for $V_1$ might actually run in the manager nodes in concrete realizations.
\subsection{BFT remote clock reading}
Firstly, to be compatible with the underlying protocol $\mathcal{P}$, we give the definition of the initially $\delta$-synchronized state.
\begin{definition}
\label{def_synchronized}
$\mathcal{L}$ is initially $\delta$-synchronized upon $G$ with $\mathcal{P}$ at $t$ iff $t$ is not in any updating span of $C_i$ for all $i\in U$ and
\begin{eqnarray}
\label{eq:def_precision} t\geqslant t_0+\Delta_{0} \land \forall i,j\in U: \mathring{d}(C_i(t),C_j(t))\leqslant \delta
\end{eqnarray}
\end{definition}

Now suppose that the system $\mathcal{L}$ is initially $\delta_{\mathtt{I}}$-synchronized (upon $G$ with $\mathcal{P}$, the same below) at $t_{1}$.
With this, to provide BFT-CS for the nonfaulty nodes in $G$, the most natural method is to run the $\mathcal{P}$ protocol for each pair of nodes $j\in U_1$ and $i\in U_0$ with $j$ being the server and $i$ being the client.
Then, for every $t\geqslant t_{1}$, each node $i\in U_0$ can remotely read the local clock $C_j(t)$ of $j\in U_1$ as $\hat{C}_{j,i}(t)$ in $i$ at $t$ with an error bounded by $\varepsilon_{0}$.
Now as the local clocks of the nodes in $U$ are initially synchronized within $\delta_{\mathtt{I}}$, every node $i\in U_0$ knows $\mathring{d}(\hat{C}_{j,i}(t),C_i(t))\leqslant \hat\delta(t)$ with some bounded $\hat\delta(t)$ when $j\in U_1$ and $t\in [t_{1}, t_{1}+k\delta_{0}]$ with $k\geqslant 1$ being a bounded integer.
Thus, by computing the actual difference of $C_i(t)$ and $\hat{C}_{j,i}(t)$ as $\tau_{j,i}(t)=\hat{C}_{j,i}(t)\ominus(C_i(t)\ominus\hat\delta(t))$, $i$ knows the values $\tau_{j,i}(t)$ are within a bounded range for all remote nodes $j\in U_1$.
So, by taking the median of $\tau_{j,i}(t)$ for all $j\in V_1$ in each node $i$, the returned values of the FTA (fault-tolerant averaging \citep{Dolev1986Approximate}) operations in all nodes $i\in U_0$ would be in a bounded range.
Following this simplest idea, denoting the underlying server-client $\mathcal{P}$ protocol running for the server $j$ and client $i$ as $\mathcal{P}_{j,i}$ (referred to as the forward $\mathcal{P}$ protocol), the basic BFT remote clock reading algorithm $\mathtt{BFT\_READ}$ is shown in Fig.~\ref{fig:algo_simplest}.
For simplicity, we assume that the algorithms are sequentially executed, in which a pending function (i.e., a function should but not yet be executed) in each node $i\in U$ would not be executed during the ongoing execution (if it exists) of any function in $i$.
If there are several pending functions in $i$, their execution orders can be arbitrarily scheduled as long as the overall maximal message delay is still bounded in $\delta_\mathtt{d}$.

\alglanguage{pseudocode}
\algrenewcommand{\algorithmiccomment}[1]{\hskip1em//#1}
\begin{figure}[htbp]
\centering
\begin{algorithmic}[1]
\Statex \textbf{for every node $i\in U_0$:}
\Statex \textbf{{initialize at $t$}} \Comment{with the initially $\delta_{\mathtt{I}}$-synchronized state}
    \State  run $\mathcal{P}_{j,i}$ for each $j\in V_1$;
\Statex \textbf{{readClock at $t$}} \Comment{read remote clocks at $t$}
    \State  $\tau:=C_i(t)$; determine $\hat\delta(t)$ as $\delta$;
    \ForAll {$j\in V_1$}
    ~  $\tau_{j,i}:=\hat{C}_{j,i}(t)\ominus(\tau\ominus\delta)$;
    \EndFor
    \State  set $\tau$ as the median of $\tau_{j,i}$ for all $j\in V_1$;\Comment{with $n_1>2f_1$}
    \State  $\textit{offset}_i:=\tau\ominus\delta$
\Statex
\Statex \textbf{for every node $j\in U_1$:}
\Statex \textbf{{initialize at $t$}} \Comment{with the initially $\delta_{\mathtt{I}}$-synchronized state}
    \State  run $\mathcal{P}_{j,i}$ for each $i\in V_0$;
\end{algorithmic}
\caption{The $\mathtt{BFT\_READ}$ algorithm.}\label{fig:algo_simplest}
\end{figure}

Note that the $\mathtt{BFT\_READ}$ algorithm does not require that the node $i\in U_0$ must actually \emph{adjust} its own clock with the $\mathtt{readClock}$ function.
It depends on concrete applications.
Sometimes, calling the $\mathtt{readClock}$ function in responding to some irregular local events in $i$ would suffice.
In other situations where the synchronized clocks are frequently referenced, the $\mathtt{readClock}$ function can also be called in $i$ to periodically adjust the logical clock $L_i(t)$ in tracing the synchronized clock at any given $t$.
As we allow $n_1=2f_1+1$, the median function is used to tolerant one Byzantine node in $V_1$ without the convergence property.

Obviously, the $\mathtt{BFT\_READ}$ algorithm along has several problems.
Firstly, during each call of the $\mathtt{readClock}$ function, the bound $\hat\delta(t)$ is dynamically determined.
Surely $\hat\delta(t)$ can also be always determined as a constant number.
But as the local clocks of nodes in $U_1$ would drift away from the initial synchronization precision $\delta_{\mathtt{I}}$ without further synchronization, the median taken for the circularly-valued remote clocks may not always be correct if $\hat\delta(t)$ is constant.
Secondly, the median function can only ensure its outputs in nodes of $U_0$ are within the range of the original inputs from $U_1$.
Now as the ranges of $\tau_{j,i}(t)$ for $j\in U_1$ in each $i$ would grow wider with the accumulated clock drifts in $U_1$, the worst-case synchronization error $\delta'(t)$ in $U_0$ would grow larger accordingly.
In overcoming this, the local clocks of nodes in $U_1$ should also be periodically synchronized.

\subsection{The basic synchronizer}
To synchronize the local clocks of nodes in $U_1$, here we want to simulate the synchronous approximate agreement \citep{Dolev1986Approximate} upon the CCBN $G$ with $n_0>3f_0$ and $n_1> 2f_1$.
Concretely, with the initial precision $\delta_{\mathtt{I}}$, besides running the forward $\mathcal{P}_{j,i}$ protocols as clients, the nodes in $U_0$ can also act as servers to reversely synchronize the nodes in $U_1$ with the backward $\mathcal{P}_{i,j}$ protocols.
The so-called backward $\mathcal{P}_{i,j}$ protocols are very like the ones proposed in ReversePTP.
The main difference is that there are $n_1$ nodes to be synchronized, not just the central node in ReversePTP.
Despite this difference, both the ReversePTP instances and the common PTP instances can be employed in realizing the backward $\mathcal{P}_{i,j}$ protocols.
Upon this, the basic BFT-CS algorithm (also called the \emph{basic synchronizer}) $\mathtt{BFT\_SYNC}$ is shown in Fig.~\ref{fig:algo_simplest2}.

\alglanguage{pseudocode}
\algrenewcommand{\algorithmiccomment}[1]{\hskip1em//#1}
\begin{figure}[htbp]
\centering
\begin{algorithmic}[1]
\Statex \textbf{for every node $i\in U_0$:}
\Statex \textbf{{initialize at $t$}} \Comment{with the initially $\delta_{\mathtt{I}}$-synchronized state}
    \State  run $\mathcal{P}_{i,j}$ and $\mathcal{P}_{j,i}$ for each $j\in V_1$;
    \State  $\textit{offset}_i:=0$;
    ~  reset timer $\tau_{w}$;
\Statex \textbf{{at local-time $k\tau_0+\delta_{3}$}} \Comment{read the new clock}
    \State  writeLogicalClock$(V_1,C_i(t),\delta_{6})$;\label{code:sync_read}
    \State set timer $\tau_{w}$ with $\delta_{4}$ ticks;
\Statex \textbf{{when timer $\tau_{w}$ is expired}}
    \State  $C_i(t):=C_i(t)\oplus\textit{offset}_i$; \label{code:sync_offset_update} \Comment{adjust the local clock}
    \State  $\textit{offset}_i:=0$;
\Statex \textbf{{writeLogicalClock$(R,\tau,\delta)$ at $t$}} \Comment{write the logical clock}
    \ForAll {$j\in R$}
    ~  $\tau_{j,i}:=\min\{\hat{C}_{j,i}(t)\ominus\tau\oplus\delta,2\delta\}$;\label{code:sync_compute_logical}
    \EndFor
    \State  set $\tau$ as the median of $\{\tau_{j,i}\mid j\in V_1\}$;\Comment{with $n_1>2f_1$}
    \State  $\textit{offset}_i:=\tau\ominus\delta$
\Statex
\Statex \textbf{for every node $j\in U_1$:}
\Statex \textbf{{initialize at $t$}} \Comment{with the initially $\delta_{\mathtt{I}}$-synchronized state}
    \State  run $\mathcal{P}_{j,i}$ and $\mathcal{P}_{i,j}$ for each $i\in V_0$;
    \State  $\textit{offset}_j:=0$;
    ~  reset timer $\tau_{w}$;
\Statex \textbf{{at local-time $k\tau_0+\delta_{1}$}} \Comment{read the new clock}
    \State  writeLogicalClock$(V_0,C_j(t),\delta_{5})$;\label{code:backward_sync_read}
    \State set timer $\tau_{w}$ with $\delta_{2}$ ticks;
\Statex \textbf{{when timer $\tau_{w}$ is expired}}
    \State  $C_j(t):=C_j(t)\oplus\textit{offset}_j$; \label{code:backward_sync_offset_update}  \Comment{adjust the local clock}
    \State  $\textit{offset}_j=0$;
\Statex \textbf{{writeLogicalClock$(R,\tau,\delta)$ at $t$}} \Comment{write the logical clock}
    \ForAll {$i\in V_0$}
    \If {$i\in R$}
      $\tau_{i,j}:=\min\{\hat{C}_{i,j}(t)\ominus\tau\oplus\delta,2\delta\}$;\label{code:backward_sync_compute_logical}
    \Else
    ~  $\tau_{i,j}:=0$;
    \EndIf
    \EndFor
    \State  set $\tau_1$ and $\tau_2$ as the $(f_0+1)$th smallest and largest $\tau_{i,j}$;\label{code:backward_sync_fta}
    \State  $\textit{offset}_j:=((\tau_1+\tau_2)/2)\ominus\delta$ \Comment{FTA with $n_0>3f_0$}\label{code:backward_sync_fta2}
\end{algorithmic}
\caption{The $\mathtt{BFT\_SYNC}$ algorithm.}\label{fig:algo_simplest2}
\end{figure}

During the initialization of the basic synchronizer, every nonfaulty node runs both the forward and backward $\mathcal{P}$ instances and resets its logical clocks and timers.
Here we say a timer (such as the timer $\tau_{w}$) is reset (denoted as $\tau_{w}=\tau_{max}$) if it is closed and would not run again before the next scheduling of it.
And we say a timer is set with $\delta$ if it is scheduled with a timeout $\delta$ after which the timer would be expired and reset.
The timeout is counted with the ticks of the hardware clock in case it is affected by upper-layer clock adjustments.
For clarity, all \emph{ticks} referred to in this paper are the ticks of the hardware clocks.
With this, for every $i\in U_0$, at each local-time $k\tau_0+\delta_{3}$ (for $k\in [[\tau_{max}/\tau_0]]$ with $\tau_{max}\bmod \tau_0=0$), $i$ reads the remote clocks and use the median of these readings as the logical clock of $i$.
After another $\delta_{4}$ ticks, $i$ adjusts its local clock with its logical clock.
Similarly, for the backward synchronization, each node $j\in U_1$ reads the remote clocks and uses the fault-tolerant averaging \citep{Dolev1986Approximate} of these readings as its logical clock at each local-time $k\tau_0+\delta_{1}$.
Then, $j$ uses it to adjust its local clock after another $\delta_{2}$ ticks.

Note that in line \ref{code:sync_offset_update} and line \ref{code:backward_sync_offset_update}, we allow $C_i(t)$ and $C_j(t)$ to be adjusted by $L_i(t)$ and $L_j(t)$, respectively.
This is necessary as the underlying $\mathcal{P}$ protocol in the server nodes should use the adjusted clocks rather than the original freely-drifting ones to ensure the differences of the referenced clocks in all nonfaulty server nodes being always in a bounded range.
But to avoid undesired asynchronous clock adjustments, firstly, the newly acquired clock values are not directly written to the local clocks.
Instead, the new values are first written to the logical clocks (with lines \ref{code:sync_read} and \ref{code:backward_sync_read}) and then written to the local clocks after some statically determined delays.
This is for simulating the synchronous approximate agreement \citep{Dolev1986Approximate} upon $G$ with lines \ref{code:backward_sync_fta} and \ref{code:backward_sync_fta2}.
And secondly, in lines \ref{code:sync_compute_logical} and \ref{code:backward_sync_compute_logical}), the offsets of the logical clocks would always be within $[0,2\delta]$.
With this, the adjustments of the local clocks would be no more than $\delta_\mathtt{I}$ even when the system is not initially $\delta_{\mathtt{I}}$-synchronized.
As the clock adjustments performed by the basic synchronizer are for maintaining some synchronized states of the system, these clock adjustments are called the basic adjustments.

\begin{figure}[htbp]
\centerline{\includegraphics[width=1.8in]{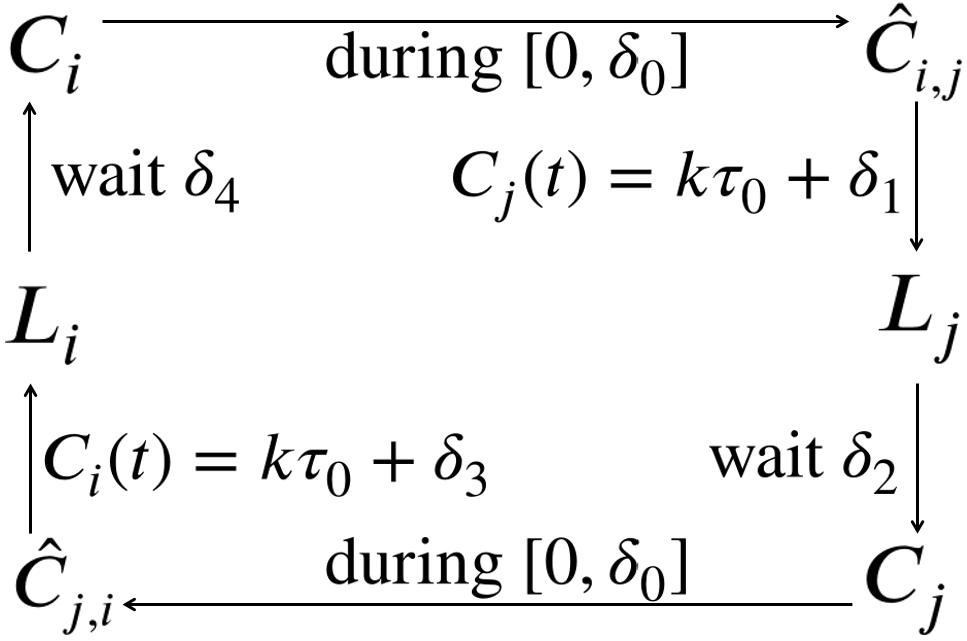}}
\caption{The temporal dependencies of the clocks.}
\label{fig:clock_dependency}
\end{figure}

In Fig.~\ref{fig:clock_dependency}, the temporal dependencies of the referred clocks are described with the labeled arrows.
The clocks $C_i$, $L_i$ and $\hat{C}_{j,i}$ (on the left side of Fig.~\ref{fig:clock_dependency}) are of the node $i\in U_0$.
And the clocks $C_j$, $L_j$ and $\hat{C}_{i,j}$ (on the right side of Fig.~\ref{fig:clock_dependency}) are of the node $j\in U_1$.
For the forward synchronization, when $C_i(t)=k\tau_0+\delta_{3}$ is satisfied, $L_i$ would be written in the $\mathtt{writeLogicalClock}$ function in $i$ with the remote clock readings $\hat{C}_{j,i}$ from all $j\in V_1$.
Then, $C_i$ would be written with $L_i$ after $\delta_{4}$ ticks in $i$.
And then, for the backward synchronization, with the underlying $\mathcal{P}_{i,j}$ protocol, $\hat{C}_{i,j}$ can be updated with the adjusted $C_i$ during the next $\delta_{0}$ time (here we assume that the actual delay can be arbitrarily distributed in $[0,\delta_{0}]$).
So by properly setting $\delta_{1}$, $L_j$ can be correctly written with the all updated $\hat{C}_{i,j}$ for all $i\in U_0$.
And by waiting for another $\delta_{2}$ ticks, $C_j$ can be correctly written with $L_j$.

\begin{figure}[htbp]
\centerline{\includegraphics[width=3.0in]{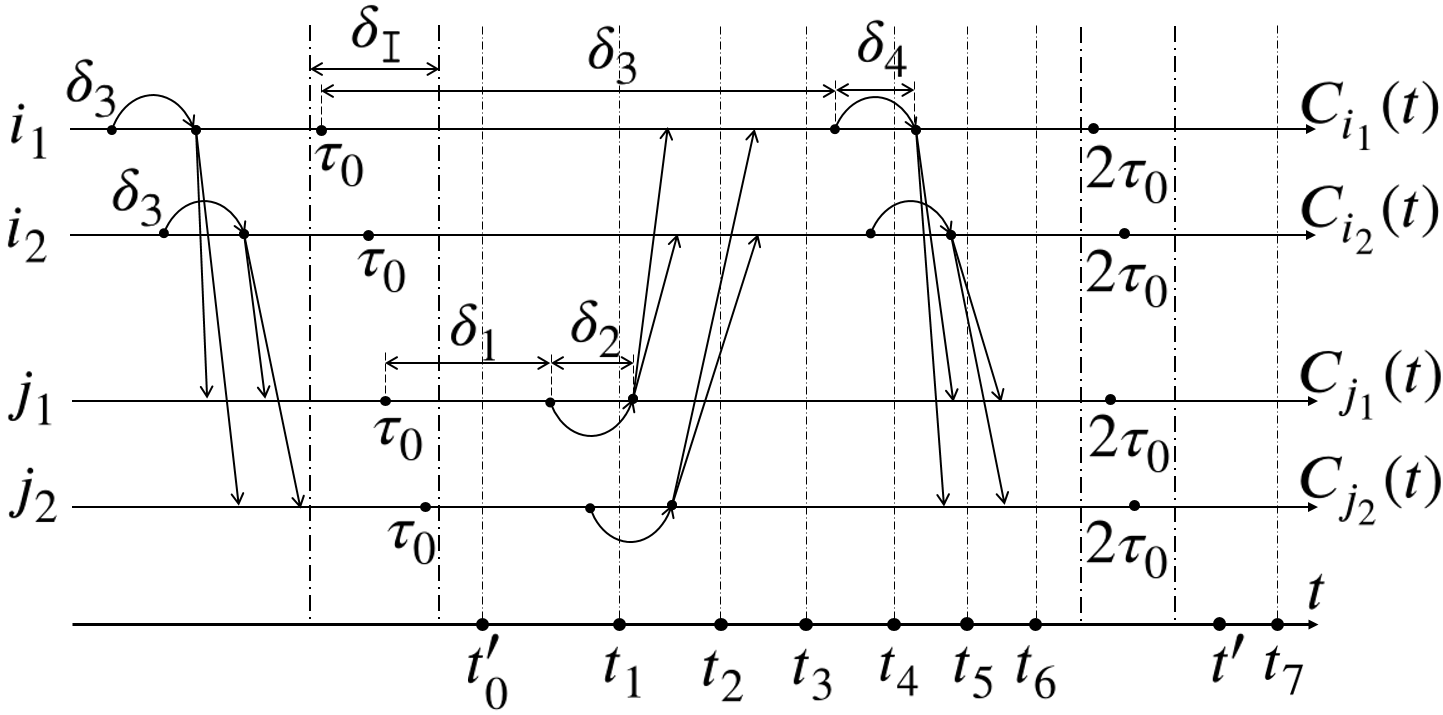}}
\caption{The strictly separated synchronization phases.}
\label{fig:basicprocess}
\end{figure}

So, the remaining problem is to determine the time parameters $\delta_{1}$, $\delta_{2}$, $\delta_{3}$, $\delta_{4}$, $\delta_{5}$, and $\delta_{6}$.
Firstly, $\mathring{d}(\hat{C}_{i,j}(t),C_j(t))\leqslant\delta_{5}$ and $\mathring{d}(\hat{C}_{j,i}(t),C_i(t))\leqslant\delta_{6}$ should hold in executing line~\ref{code:sync_read} and line~\ref{code:backward_sync_read} of the $\mathtt{BFT\_SYNC}$ algorithm with the initially $\delta_{\mathtt{I}}$-synchronized state.
Secondly, $\delta_{1}$, $\delta_{2}$, $\delta_{3}$, and $\delta_{4}$ should be determined to ensure the basic synchronization procedure simulating the desired synchronous approximate agreement, as is shown in Fig.~\ref{fig:basicprocess}.

%
%$i_1\in U_0$ $C_{i_1}(t)$
%
%$i_2\in U_0$ $C_{i_2}(t)$
%
%$j_1\in U_1$ $C_{j_1}(t)$
%
%$j_2\in U_1$ $C_{j_2}(t)$
%
%$k\tau_0$
%
%$2\tau_0+\delta_{1}+\delta_{2}$
%$2\tau_0+\delta_{3}+\delta_{4}$
%
%$(k+1)\tau_0$
%
%$C_{j}(t)=k\tau_0+\delta_{1}$
%$C_{i}(t)=k\tau_0+\delta_{3}$
%wait $\delta_{2}$
%wait $\delta_{4}$
%during $[0,\delta_{0}]$

In Fig.~\ref{fig:basicprocess}, the fastest and slowest nodes in $U_0$ ($U_1$) are denoted as $i_1$ and $i_2$ ($j_1$ and $j_2$), respectively.
It should be noted that the actual slowest and fastest nodes can change over time.
Here the case is just for describing the desired basic synchronization procedure.
Arrows still represent the influences between the clocks.
For example, the leftmost curved arrow (on the local-time of $i_1$) represents that the local clock $C_i$ is adjusted at $\delta_{3}+\delta_{4}$ with the logical clock $L_i$ being written at $\delta_{3}$.
And the straight arrows represent the clock distributions from the server-clocks to the client-clocks with the underlying $\mathcal{P}$ protocols.
Here, as the local clocks of all nodes in $U$ are initially synchronized within $\delta_{\mathtt{I}}$, the synchronization phases (separated by the long dotted lines in Fig.~\ref{fig:basicprocess}) in the distributed nonfaulty nodes can be well-separated in real-time if the synchronization precision can be maintained within some fixed bounds.
In Section~\ref{sec:Analysis}, we would see that with properly configured time parameters, the desired synchronization precision can be maintained in $\mathcal{L}$ with the initially $\delta_{\mathtt{I}}$-synchronized state.
Note that the basic settings of the time parameters are for strictly separating the synchronization phases shown in Fig.~\ref{fig:basicprocess}.
Actually, by setting one or both of the parameters $\delta_{4}$ and $\delta_{2}$ being $0$, the synchronization procedure can also be realized in a rather wait-free manner.
For simplicity, we take strictly separated synchronization phases in this paper.
With this, a basic synchronization round of the initially $\delta_{\mathtt{I}}$-synchronized $\mathcal{L}$ can be defined with any periodically appearing synchronization phase shown in Fig.~\ref{fig:basicprocess}.
For instance, the time interval $[t_0',t']$ can be viewed as a basic synchronization round of $\mathcal{L}$.
And for every $i\in U$, when $C_i(t)\in [(k-1)\tau_0,k\tau_0)$ with $k\geqslant 1$, we say $i$ is in its $k$th local basic synchronization round.

\subsection{The strong synchronizer}
Besides the \emph{basic synchronizer}, an additional \emph{pulse synchronizer} is provided with the $\mathtt{BFT\_PULSE\_SYNC}$ algorithm, as is shown in Fig.~\ref{fig:algo_pulse}.
The \emph{basic synchronizer} together with the \emph{pulse synchronizer} are called the \emph{strong synchronizer}.
Here the readers might wonder why more than one synchronization algorithm is provided.
Roughly speaking, with the strong synchronizer, we can provide some easier evidence that once such evidence is observed in a nonfaulty node $j\in V_1$, $j$ would know that the system would be stabilized in an expected way.
Upon this, if all nodes in $U_1$ observe such evidence for a sufficiently long time, the extra self-stabilizing procedure would not be performed.
We would further explain this when we construct the stabilizer with this strong synchronizer.
Here we first describe the $\mathtt{BFT\_PULSE\_SYNC}$ algorithm and its relationship to the $\mathtt{BFT\_SYNC}$ algorithm.
For simplicity, we assume $n_0>5f_0$ for the $\mathtt{BFT\_PULSE\_SYNC}$ algorithm.

\alglanguage{pseudocode}
\algrenewcommand{\algorithmiccomment}[1]{\hskip1em//#1}
\begin{figure}[htbp]
\centering
\begin{algorithmic}[1]
\Statex \textbf{for every node $i\in U_0$:}
\Statex \textbf{{at local-time $kk_\mathtt{pls}\tau_0$}}
    \If {$\delta_{15}$ ticks passed since the last pulsing event}\label{code:pulse_sync_pulse_con}
    \State  send pulse-$k$ to each $j\in V_1$;\label{code:pulse_sync_pulse_send0}\Comment{the pulsing event}
    \EndIf
\Statex \textbf{{always}}
    \State  $P_k=\{j\mid i$ receives pulse-$k$ from $j$ in the latest $\delta_{16}$ ticks$\}$; \label{code:pulse_sync_always_receive0}
    \If {$\exists k':|P_{k'}|\geqslant n_1-f_1$}\Comment{with $n_1>2f_1$} \label{code:pulse_sync_preempitve_con0}
    \State  $k^*:=k'$;\label{code:pulse_sync_preempitve_start0}
    ~  set timer $\tau_{w}^*$ with $\delta_{17}$ ticks;
    \EndIf
\Statex \textbf{{when timer $\tau_{w}^*$ is expired}}
    \State  ${\tau'}:=k^*k_\mathtt{pls}\tau_0+\delta_{9}$;
    \ForAll {$i\in V_0$}
    $\tau_{j,i}'=\hat{C}_{j,i}(t)\ominus{\tau'}$; \label{code:pulse_sync_input0}
    \EndFor
    \State  set $\tau$ as the median of $\{\tau_{j,i}'\mid j\in V_1\}$;\Comment{with $n_1>2f_1$}
    \State  $C_i(t):={\tau'}\oplus\tau$;\label{code:pulse_sync_median}
    ~  $\textit{offset}_i=0$;\label{code:pulse_sync_preemptive_do0}
    \State  set $\textit{protect}_i$ as $1$ until $C_i\bmod \tau_0=0$;\label{code:pulse_sync_preemptive_stop0}
\Statex
\Statex \textbf{for every node $j\in U_1$:}
\Statex \textbf{{always}}
    \State  $P_k=\{i\mid j$ receives pulse-$k$ from $i$ in the latest $\delta_{10}$ ticks$\}$; \label{code:pulse_sync_always_receive}
    \If {$\exists k':|P_{k'}|\geqslant n_0-2f_0$}\Comment{with $n_0>5f_0$} \label{code:pulse_sync_preempitve_con}
    \State  $k^*:=k'$;\label{code:pulse_sync_preempitve_start}
    \State  set timer $\tau_{w}^*$ with $\delta_{11}$ ticks;
    \EndIf
\Statex \textbf{{when timer $\tau_{w}^*$ is expired}}
    \State  ${\tau'}:=k^*k_\mathtt{pls}\tau_0+\delta_{8}$;
    \ForAll {$i\in V_0$}
    $\tau:=\hat{C}_{i,j}(t)\ominus\tau'$;\label{code:pulse_sync_raw_input}
    \If {$\tau\leqslant \delta_{7}$}
    ~  $\tau_{i,j}'=\tau$; \label{code:pulse_sync_input}
    \Else
    ~  $\tau_{i,j}':=0$;
    \EndIf
    \EndFor
    \State  set $\tau_1'$ and $\tau_2'$ as the $(f_0+1)$th smallest and largest $\tau_{i,j}'$;
    \State  $C_j(t):={\tau'}\oplus(\tau_1'+\tau_2')/2$;~  $\textit{offset}_j=0$;\Comment{FTA}\label{code:pulse_sync_preemptive_do}
    \State  set $\textit{protect}_j$ as $1$ until $C_j\bmod \tau_0=0$;\label{code:pulse_sync_preemptive_stop}
\Statex \textbf{{at local-time $kk_\mathtt{pls}\tau_0+\delta_{12}$}}
    \If {timer $\tau_{w}^*$ is set in the last $\delta_{12}$ ticks}
    \State  send pulse-$k^*$ to each $i\in V_0$;\label{code:pulse_sync_pulse_send}
    \EndIf
\end{algorithmic}
\caption{The $\mathtt{BFT\_PULSE\_SYNC}$ algorithm.}\label{fig:algo_pulse}
\end{figure}

\begin{figure}[htbp]
\centerline{\includegraphics[width=3.3in]{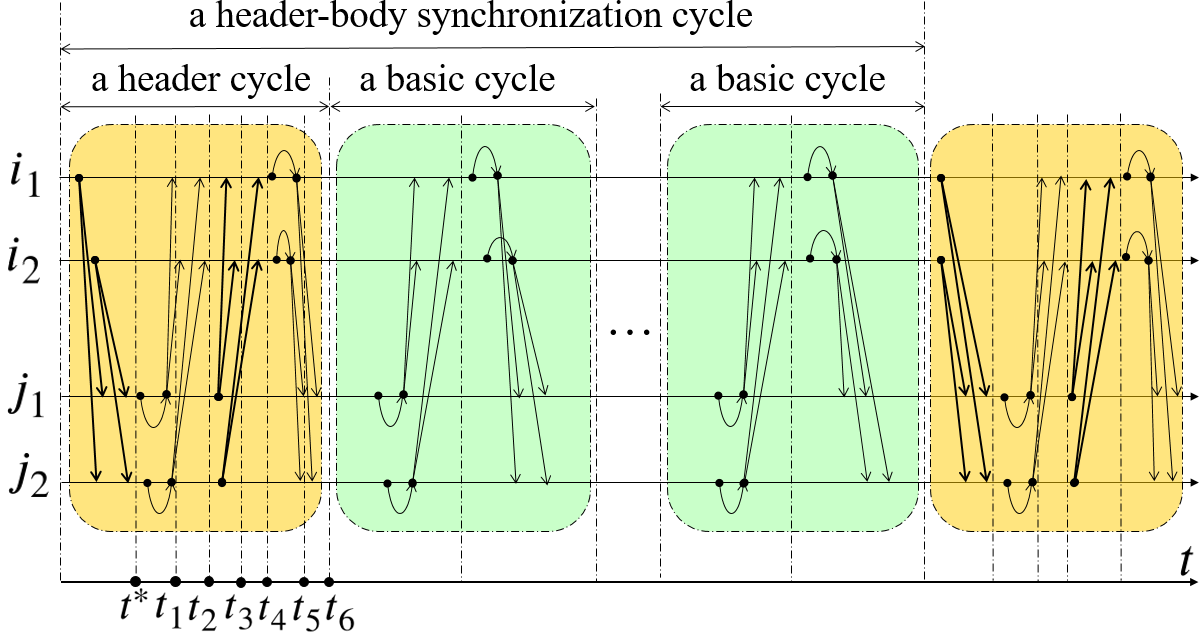}}
\caption{The desired header-body synchronization procedure.}
\label{fig:headerbody}
\end{figure}

As is shown in Fig.~\ref{fig:algo_pulse}, firstly, by setting $k_\mathtt{pls}> 1$, the additional synchronization would be performed with a lower frequency than the basic synchronization.
This is for well separating the additional pulse-like sparse synchronization events.
Besides, with line~\ref{code:pulse_sync_pulse_con} of the $\mathtt{BFT\_PULSE\_SYNC}$ algorithm ($i$ would count the ticks since the beginning if $i$ has not yet sent the first pulse), a node $i\in U_0$ would not send any two pulses within $\delta_{15}$ ticks.
This would provide some good properties for constructing the overall IS-BFT-CS solution.
Here, when it runs in the desired way, this additional synchronization procedure adds some header rounds (or saying \emph{headers}) into the original synchronization procedure, as is shown in Fig.~\ref{fig:headerbody}.
In each such synchronization header (the yellow block in Fig.~\ref{fig:headerbody}), there should be at least $n_0-2f_0$ nodes in $U_0$ sending their pulses (shown in Fig.~\ref{fig:headerbody} as bold arrows) in a short duration no more wider than $\delta_{10}/(1+\rho)-\delta_\mathtt{d}$.
In this sense, these nodes in $U_0$ are called a pulsing clique in $U_0$, as all their pulses are within a sufficiently narrow duration.

Then, to perform the desired additional synchronization in the presence of such a pulsing clique, the lines from \ref{code:pulse_sync_always_receive} to \ref{code:pulse_sync_preemptive_stop} of the $\mathtt{BFT\_PULSE\_SYNC}$ algorithm (denoted as the $\mathtt{B_1}$ block) should be executed with a higher priority than all lines of the $\mathtt{BFT\_SYNC}$ algorithm.
Namely, when a node $j\in U_1$ writes the logical clock $L_j$ and adjusts the local clock $C_j$ in executing the $\mathtt{B_1}$ block, any attempt to write $L_j$ or $C_j$ in the $\mathtt{BFT\_SYNC}$ algorithm would be preempted and canceled during a bounded time interval.
This is for avoiding the undesired output of the FTA operation of the $\mathtt{BFT\_SYNC}$ algorithm to overwrite the desired output of the $\mathtt{B_1}$ block in the presence of a desired pulsing clique.
Moreover, we use the a flag $\textit{protect}_j\in\{0,1\}$ (with the default value $0$) in the algorithm for each node $j\in U$ to indicate that if the clocks of $j$ should be protected from being adjusted outside this algorithm.
Concretely, the clocks of $j$ can be adjusted outside the algorithm if and only if $\textit{protect}_j$ is $0$ at $t$ in node $j$.
Besides, the executions of the $\mathtt{B_1}$ block can also be preempted and canceled by themselves when two or more such executions are temporally overlapped.
In other words, the latter execution of the $\mathtt{B_1}$ block always has the higher priority (with even canceling the cancelation of clock-writings implemented in the former executions).
Thus, with a pulsing clique, the local clocks of all $j\in U_1$ would be semi-synchronously adjusted with the line \ref{code:pulse_sync_preemptive_do} of the $\mathtt{BFT\_PULSE\_SYNC}$ algorithm.
And all these local clocks would at least be synchronized with the precision in the same order of $\delta_{10}$.
Similarly, the nodes in $U_0$ would also be synchronized with such a coarse precision in the presence of the desired pulsing clique.
Then, although this precision could be coarser than the desired final synchronization precision, it is not a problem as the synchronization header is followed by the synchronization body in executing the $\mathtt{BFT\_SYNC}$ algorithm.
Namely, in the synchronization body (shown in Fig.~\ref{fig:headerbody} as the green blocks following the leftmost yellow one), a $(k_\mathtt{pls}-1)$-round synchronous approximate agreement is simulated (one such round is also shown in Fig.~\ref{fig:basicprocess} in $[t_0',t']$).
With this, by the end of the synchronization body of the desired synchronization procedure, the local clocks (and also the logical clocks) of all nodes in $U$ would be synchronized with the desired precision.
In simulating the approximate agreement, the convergence rate can be further improved by employing the advanced FTA functions given in \cite{Dolev1986Approximate}.
Here the basic solution employs the basic FTA function (with convergence rate $1/2$) for simplicity.

Generally, this header-body synchronization procedure is called a two-stage synchronization procedure.
To make this two-stage synchronization procedure work in the presence of a pulsing clique in $U_0$, firstly, the first stage should deterministically bring all local clocks of the nodes in $U_1$ into the expected coarser precision.
This is implemented by the $\mathtt{BFT\_PULSE\_SYNC}$ algorithm by making all pulsing cliques in $U_0$ being well-separated in real-time.
Secondly, the second stage should deterministically simulate the synchronous approximate agreement.
This is implemented by the $\mathtt{BFT\_SYNC}$ algorithm with an initially $\delta_{\mathtt{I}}$-synchronized state at the end of the first stage.
In Section~\ref{sec:Analysis}, we would show that this procedure can be performed with properly configured time parameters.
For simplicity, the header cycle (i.e., the nominal duration of a header) is also set as the basic cycle $\tau_0$ (i.e., the nominal duration of basic synchronization round).
And a header can be viewed as a special kind of basic synchronization round.

\section{Basic IS-BFT-CS solution}
\label{sec:SSBFTCS}
The $\mathtt{BFT\_SYNC}$ and $\mathtt{BFT\_PULSE\_SYNC}$ algorithms are not self-stabilizing, since either an initially $\delta_{\mathtt{I}}$-synchronized state or a pulsing clique is required in executing these algorithms.
For stabilization, the system should be synchronized in some desired time with all possible initial states.
In this section, we provide a basic IS-BFT-CS solution.
\subsection{The problem of stabilization}
As there might be no initially $\delta_{\mathtt{I}}$-synchronized state at $t_{0}$ nor any desired pulsing clique since $t_0$, reliable synchronization cannot be established with only the strong but still non-stabilizing synchronizer.
For stabilization, some kind of BFT stabilizers can be employed.
The so-called BFT stabilizers, such as the ones proposed and utilized in \cite{Daliot2005SSBProtocols,Daliot2006Agreement,Dolev2007stabilizing}, are able to convert non-stabilizing BFT protocols to the corresponding stabilizing ones.
For example, the self-stabilizing DBA (SS-DBA) algorithm proposed in \cite{Daliot2006Agreement} is used as a primitive in construction the deterministic SS-BFT-CS in \cite{DolevPulseBoundedDelay2007}.
For some other examples, some resynchronization algorithms are used as BFT stabilizers in the SS-BFT-CS algorithms provided in \cite{Lenzen2019AlmostConsensus,Dolev2014PulseGeneration}.
Obviously, if the manager nodes are fully connected, we can directly employ some existing SS-BFT-CS algorithms \citep{DolevPulseBoundedDelay2007,Dolev2014PulseGeneration,Khanchandani2016Optimal,Lenzen2019AlmostConsensus} to construct the core synchronization system and then distribute the clocks of the manager nodes to the whole system.

However, the existing SS-BFT-CS solutions have several disadvantages in the specific context of IoT networks.
Firstly, building upon the classical bounded-delay assumption, all the existing SS-BFT-CS solutions are with synchronization precision no better than $o(d')$, where $d'$ is the maximal message delay in the corresponding communication networks.
In contrast, CS protocols such as PTP can often achieve better precision with several low-cost hardware and software optimizations.
Secondly, most existing SS-BFT-CS solutions are constructed by periodically executing some kind of BA protocol, which generates additional complexity even when the system is stabilized.
In contrast, CS protocols such as PTP require very sparse resources in maintaining the stabilized state of the system.
Thirdly, although some randomized SS-BFT-CS algorithm does not rely on BA protocols, the expectation of the stabilization time is at least $O(n)$, where $n$ is the number of the synchronization nodes in the system.
In contrast, CS protocols such as PTP trivially have a deterministic constant stabilization time.
Fourthly, almost all existing SS-BFT-CS solutions require CCN in exchanging the synchronization messages.
In contrast, the most common PTP protocols can run upon tree topologies without the message being exchanged between client nodes (with a pre-configured grandmaster).
And lastly, migrating the SS-DBA-based BFT-stabilizer into CCBN is also not a trivial task and would generate many more messages, especially with $n_1=2f_1+1$.
In contrast, some variants of PTP, such as ReversePTP, do not require exchanging any message between the clients in electing a new grandmaster.

\begin{figure}[htbp]
\centerline{\includegraphics[width=2.9in]{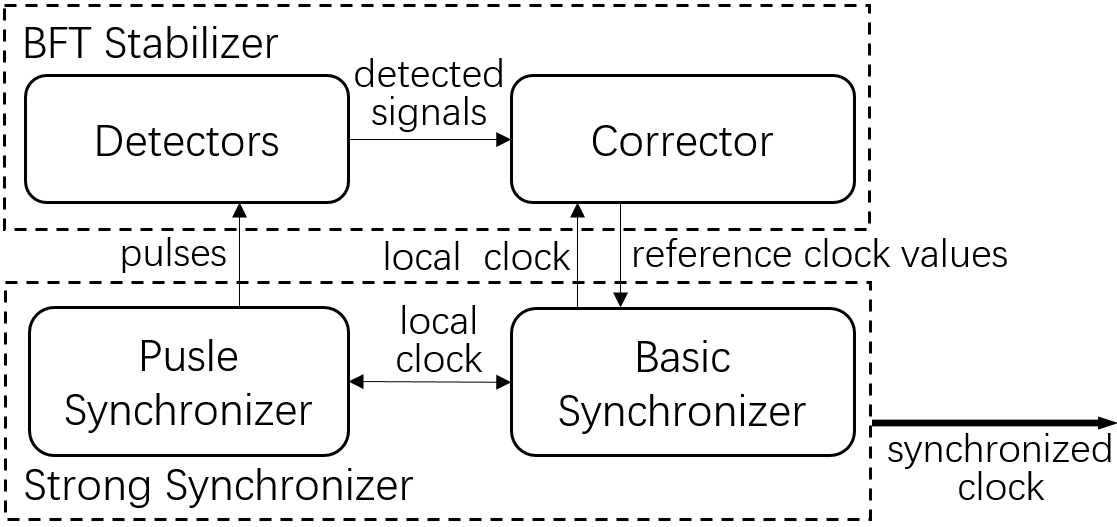}}
\caption{The overall framework of IS-BFT-CS.}
\label{fig:framework}
\end{figure}

In mitigating these disadvantages, we provide a basic IS-BFT-CS solution upon CCBN with discreetly utilized external times.
Generally, the overall framework of the IS-BFT-CS solutions is shown in Fig.~\ref{fig:framework}.
With the developed strong synchronizer, the main problem here is to construct the BFT stabilizer.

The BFT stabilizer should have the following properties.
Firstly, the system should reach the stabilized state from an arbitrary initial state in the desired time.
And secondly, for efficiency, once the system is stabilized, no nonfaulty node would detect the undesired system state.
So no corrector would be called further.
For this, the BFT stabilizer can be constructed in two steps.
In the first step, we would construct some detector (or saying state-checker, monitor, etc.) to detect some undesired state of the system.
In the second step, some corrector (or saying state-resetter, repairer) would be called to bring the state of the system into the desired one.
As is required, no single point of failure is allowed.
So the detector and corrector can only be implemented in a fault-tolerant way.

Also, here we want to design the synchronizers, detectors, and correctors in a more decoupled way.
One benefit of this would be that the basic building blocks would then be integrated with other extra supports such as external times and other resources more easily.
And with the decoupled strong synchronizer and corrector, once the system is stabilized, the corrector would not be active before the happening of the next transient system-wide failure.
With this, we expected that the synchronization precision and overall performance could be further improved.

\begin{figure}[htbp]
\centerline{\includegraphics[width=2.5in]{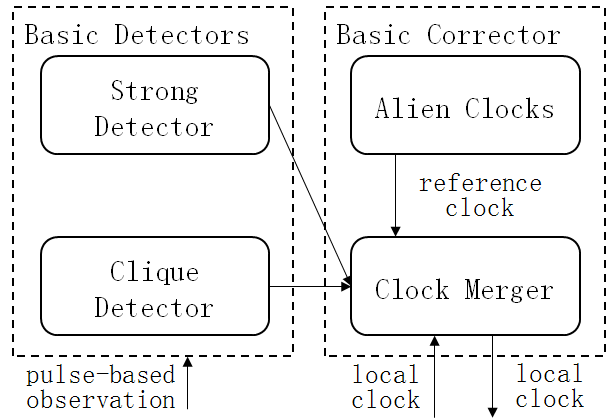}}
\caption{The basic BFT stabilizer.}
\label{fig:stabilizer}
\end{figure}

For this, the basic BFT stabilizer is built with the structure shown in Fig.~\ref{fig:stabilizer}.

\subsection{The basic detectors}
To construct the BFT stabilizer, firstly, the $\mathtt{S\_DETECTOR}$ algorithm is provided in Fig.~\ref{fig:algo_sdetector} to act as the \emph{strong} detector.
For a concrete example, the \emph{set} operation and the \emph{expired} condition of the timer $\tau_{d}$ are implemented by the line~\ref{code:detector_timer_set} and line~\ref{code:detector_timer_expired} of the $\mathtt{S\_DETECTOR}$ algorithm, respectively.
Other timers (such as $\tau_{w}^*$ and $\tau_{w}$) can be implemented similarly for fast stabilization.
Here the timer $\tau_{d}$ is used as a watchdog timer to count the ticks passed since the last satisfaction of the condition in line~\ref{code:detector_satis} of the $\mathtt{S\_DETECTOR}$ algorithm.
So if $\tau_{d}$ is expired in $j\in U_1$, $j$ knows that the system is not stabilized, by which we say $j$ is alerted.

\alglanguage{pseudocode}
\algrenewcommand{\algorithmiccomment}[1]{\hskip1em//#1}
\begin{figure}[htbp]
\centering
\begin{algorithmic}[1]
\Statex \textbf{for every node $j\in U_1$:}
~ \textbf{{always}}
    \State  $P_k=\{i\mid j$ receives pulse-$k$ from $i$ in the latest $\delta_{13}$ ticks$\}$;
    \If {$\exists {k'}:|P_{k'}|\geqslant n_0-f_0 \land $} \label{code:detector_satis}
    \State  $\tau_{d}:=H_j(t)\oplus(\delta_{14}-1)$; \label{code:detector_timer_set}
    \EndIf
    \If {$\tau_{d}\ominus H_j(t)>\delta_{14} \land \tau_{d}\neq \tau_{max}$} \label{code:detector_timer_expired}
    $\tau_{d}:=\tau_{max}$;
    \EndIf
    \State $\textit{alerted}_j:=(\tau_{d}= \tau_{max})$;
\end{algorithmic}
\caption{The $\mathtt{S\_DETECTOR}$ algorithm.}\label{fig:algo_sdetector}
\end{figure}

The detector is called \emph{strong} (a slight abuse of the concept proposed in \citep{Chandra1996Detectors}) in that all possible undesired system states would be eventually detected in all nonfaulty nodes while some of the detected ones may not be actually the undesired cases.
This kind of \emph{false alarm} is largely inevitable in designing the detector in the presence of Byzantine nodes.
But if the system is stabilized for a sufficiently long time, no false alarm would be generated.
So it leaves for some kind of \emph{correctors} to take appropriate actions in responding to the alarms (including the false-alarms) being generated in the \emph{strong} detector.

Similarly, other detectors can be designed to detect any other observable system states.
For example, the clique detector $\mathtt{Q\_DETECTOR}$ shown in Fig.~\ref{fig:algo_wdetector} can tell if there is a possible pulsing clique in the current local basic synchronization round (the line~\ref{code:pulse_sync_preemptive_stop0} and line~\ref{code:pulse_sync_preemptive_stop} of the $\mathtt{BFT\_PULSE\_SYNC}$ algorithm can be realized similarly).
The $\mathtt{S\_DETECTOR}$ and $\mathtt{Q\_DETECTOR}$ are called the basic detectors.

\alglanguage{pseudocode}
\algrenewcommand{\algorithmiccomment}[1]{\hskip1em//#1}
\begin{figure}[htbp]
\centering
\begin{algorithmic}[1]
\Statex \textbf{for every node $j\in U_1$:}
~ \textbf{{always}}
    \If {$\tau_{w}^*\neq\tau_{max}$} \label{code:detector_satis2}
      $\textit{pulsed}_j:=1$;
    \EndIf
    \If {$C_j(t)\bmod k_\mathtt{pls}\tau_0>\tau_0+(1+\rho)\delta_\mathtt{d}$}
      $\textit{pulsed}_j:=0$;
    \EndIf
\end{algorithmic}
\caption{The $\mathtt{Q\_DETECTOR}$ algorithm.}\label{fig:algo_wdetector}
\end{figure}

\subsection{The basic corrector}

The basic corrector is constructed as the $\mathtt{H\_CORRECTOR}$ algorithm shown in Fig.~\ref{fig:algo_basic_corrector} with the alien clocks (shown in the grey color in Fig.~\ref{fig:stabilizer} and given in Section~\ref{sec:Model}).
Concretely, the $\mathtt{H\_CORRECTOR}$ algorithm running in every $j\in U_1$ uses the alien clock $Y_j$ (in executing line~\ref{code:corrector_correction} of the $\mathtt{H\_CORRECTOR}$ algorithm) as some kind of temporary synchronized clock to adjust $C_j$ when the system is not stabilized.
So, when the system is not stabilized, the alien clocks $Y_j$ for all $j\in U_1$ are assumed to be at least coarsely synchronized.

\alglanguage{pseudocode}
\algrenewcommand{\algorithmiccomment}[1]{\hskip1em//#1}
\begin{figure}[htbp]
\centering
\begin{algorithmic}[1]
\Statex \textbf{for every node $j\in U_1$:}
\Statex \textbf{{at local-time $(kk_\mathtt{pls}+1)\tau_0$}}
    \State  $\textit{coin}_j:=\textit{random }(\{0,1\})$;  \label{code:corrector_coin_toss}
    \If {$\textit{EoR }$}\Comment{EoR is observed}\label{code:corrector_eor}
    \State  set $\textit{protect}_j$ as $1$; \label{code:corrector_correction_begin}
    \State  $C_j(t):=Y_j(t)$;  \label{code:corrector_correction}
    ~  $\textit{offset}_j=0$;\label{code:corrector_correction_end}
    \Else
    ~  set $\textit{protect}_j$ as $0$;
    \EndIf
\end{algorithmic}
\caption{The $\mathtt{H\_CORRECTOR}$ algorithm.}\label{fig:algo_basic_corrector}
\end{figure}

Generally, in the $\mathtt{H\_CORRECTOR}$ algorithm, not only the alien clocks but any kind of synchronized clocks can be employed to provide the reference clocks $Y_j$ for all $j\in U_1$, providing that these clocks can be coarsely synchronized when $\mathcal{L}$ is not synchronized.
However, as the stabilization time and message complexity of traditional SS-BFT-CS solutions are often prohibitively high, the alien clocks can be utilized here to reduce the stabilization time of the overall IS-BFT-CS system.
Now providing that $Y_j$ are coarsely synchronized for all $j\in U_1$ with the precision $e_1$, the $\mathtt{H\_CORRECTOR}$ algorithm would mainly act as some kind of \emph{clock merger} to merge $C_j$ and $Y_j$ at some appropriate instants.
Concretely, only if the node $j$ observes the evidence of resynchronization (EoR), $j$ would use its alien-time $Y_j(t)$ to overwrite its local-time $C_j(t)$.
To our aim, the EoR condition checked in line~\ref{code:corrector_eor} (of the $\mathtt{H\_CORRECTOR}$ algorithm, the same below) can be configured in $j$ as
\begin{eqnarray}
\label{eq:def_eor} \textit{EoR}=(\textit{alerted}_j\land (\neg\textit{pulsed}_j\lor \textit{coin}_j))
\end{eqnarray}
to give chances for both fast and stable synchronization of $C_j$ for all $j\in U_1$.
As $\neg\textit{pulsed}_j$ implies $\textit{alerted}_j$ when $\textit{EoR}$ is checked in node $j$, $\textit{EoR}$ can also be computed as $\neg\textit{pulsed}_j \lor (\textit{alerted}_j \land \textit{coin}_j) $.
Roughly speaking, in running the intro-stabilizing BFT-CS algorithms with $\textit{EoR}$, once any internal synchronizer (i.e., except the alien clocks) might work, the EoR condition is expected to be false, and thus the clock-merge operation is expected to be forbidden.
When no such internal synchronizer works, the EoR condition is expected to be true, and thus the clock-merge operation is expected to be allowed.

For this, besides the $\textit{alerted}_j$ and $\textit{pulsed}_j$ signals from the detectors, we also employ the $\textit{coin}_j$ flag, which is the result of the coin tossed in executing line~\ref{code:corrector_coin_toss} when $C_j(t)\bmod k_\mathtt{pls}\tau_0=\tau_0$.
This $\textit{coin}_j$ flag is necessary as some node in $U_1$ might observe $\textit{pulsed}_j$ while some other nodes in $U_1$ might observe $\neg\textit{pulsed}_j$ during their overlapped header-body synchronization procedures.
In this situation, there might be some nodes that want to be synchronized by the $Y$ clocks while the others do not.
To reconcile this, every node $j\in U_1$ can toss an unbiased coin during every header-body synchronization procedure to decide if it would like to be synchronized by the $Y_j$ clock or not when $\textit{alerted}_j\land \textit{pulsed}_j$ is observed.
For better performance, some biased coins can also be employed.
Here we take the unbiased coin for simplicity.
Notice that we can also compute $\textit{EoR}$ as $\textit{alerted}_j\land \textit{coin}_j$ to simplify the analysis.
However, with this simplification, some non-worst-cases optimization is also sacrificed, as it is more likely that some nodes in $U_1$ would observe $\neg\textit{pulsed}_j$ when the system is not synchronized.

Lastly, to be integrated with the strong synchronizer, the execution of the $\mathtt{H\_CORRECTOR}$ algorithm has a lower priority than that of the $\mathtt{BFT\_PULSE\_SYNC}$ algorithm.
Namely, whenever the local clock $C_j$ would be adjusted in executing the $\mathtt{BFT\_PULSE\_SYNC}$ algorithm, this adjustment would not be canceled by setting $\textit{protect}_j$ as $1$ in executing the $\mathtt{H\_CORRECTOR}$ algorithm.
Meanwhile, the execution of the $\mathtt{H\_CORRECTOR}$ algorithm still has a higher priority than that of the $\mathtt{BFT\_SYNC}$ algorithm.
It should be noted that as the $\textit{protect}_j$ flag is set as $1$ in executing the $\mathtt{BFT\_PULSE\_SYNC}$ only when $C_j\bmod k_\mathtt{pls}\tau_0\leqslant \tau_0$, the local state $\textit{protect}_j=1$ set in executing the $\mathtt{H\_CORRECTOR}$ algorithm would not be changed by executing the $\mathtt{BFT\_PULSE\_SYNC}$ algorithm.

\section{Formal analysis}
\label{sec:Analysis}
Now we show that by configuring the constant parameters referenced in the algorithms according to the constraints shown in Table~\ref{tab:parameter_relations} and Table~\ref{tab:related_relations} (some constraints are relaxed for simplicity), the provided algorithms make an IS-BFT-CS solution upon $G$.
Some concrete configurations for the constant parameters are later given in Table~\ref{tab:parameters}.

\begin{table}[htbp]
\caption{The constraints of the parameters used in the algorithms}
\label{tab:parameter_relations}
\begin{tabular}{Nl}
\hline\noalign{\smallskip}
\multicolumn{1}{c}{No.} &Constraints   \\
\noalign{\smallskip}\hline\noalign{\smallskip}
\label{re:delta1}&$\delta_{1}\geqslant (\theta_{1}+\delta_\mathtt{d})(1+\rho)$\\%
\label{re:delta2}&$\delta_{2}\geqslant \theta_{1}(1+\rho)$\\
\label{re:delta3}&$\delta_{3}\geqslant \theta_{3}(1+\rho)+\delta_{1}$\\
\label{re:delta4}&$\delta_{4}\geqslant \theta_{1}(1+\rho)+2\rho\theta_{4}$\\
\label{re:delta5}&$\delta_{5}\geqslant \delta_{\mathtt{I}}+2\rho\theta_{2}+2\varepsilon_{0}$\\
\label{re:delta6}&$\delta_{6}\geqslant \delta_{\mathtt{I}}+2\rho\theta_{4}+4\varepsilon_{0}$\\
\label{re:delta7}&$\delta_{7}\geqslant \varepsilon_{1}+(\delta_\mathtt{d}+\delta_{11}/(1-\rho)+\delta_{\mathtt{p}})(1+\rho)+\varepsilon_{0} $\\
\label{re:delta8}&$\delta_{8}= \delta_{11}(1-\rho)/(1+\rho)-\varepsilon_{0}$\\
\label{re:delta9}&$\delta_{9}= \delta_{12}+\delta_{17}(1-\rho)/(1+\rho)-\varepsilon_{0}$\\
\label{re:delta10}&$\delta_{10}\geqslant (\sigma_{7}+\delta_\mathtt{d})(1+\rho)$\\
\label{re:delta11}&$\delta_{11}\geqslant \delta_{10}+\delta_{\mathtt{p}}$\\
\label{re:delta12}&$\delta_{12}\geqslant \delta_{7}+\delta_{8}+\delta_{11}+2\delta_{\mathtt{p}}$\\
\label{re:delta13}&$\delta_{13}\geqslant \varepsilon_{1}+\delta_\mathtt{d}(1+\rho)$\\
\label{re:delta14}&$\delta_{14}\geqslant k_\mathtt{pls}(\tau_0+2\delta_{\mathtt{I}})+\delta_{\mathtt{p}}$\\
\label{re:delta15}&$\delta_{15}=k_\mathtt{pls}(\tau_0-2\delta_{\mathtt{I}})-\delta_{\mathtt{p}}\geqslant(3\sigma_{1}+\sigma_{3})(1+\rho)$\\
\label{re:delta16}&$\delta_{16}\geqslant \sigma_{11}+\delta_\mathtt{d}(1+\rho)$\\
\label{re:delta17}&$\delta_{17}\geqslant \delta_{16}+\delta_{\mathtt{p}}$\\
\label{re:tau0}&$\tau_0\geqslant \max\{\delta_{6}+(\theta_{5}+\delta_{0})(1+\rho),\delta_{12}+\delta_{\mathtt{I}}+(\sigma_{12}+\delta_{0})(1+\rho)\}$\\
\label{re:taumax}&$\tau_{max}\geqslant 4\delta_{14} \land \tau_{max}\bmod(4k_\mathtt{pls}\tau_0)=0 $\\
\label{re:kpls}&$k_\mathtt{pls}\geqslant \max\{1+\lceil \log_\alpha ((\varepsilon_{1}/2-\epsilon_\mathtt{b}/(1-\alpha))/\delta_\mathtt{I})\rceil,3\}$\\
\noalign{\smallskip}\hline
\end{tabular}
\end{table}

\begin{table}[htbp]
\caption{The other related parameters and constraints}
\label{tab:related_relations}
\begin{tabular}{Nl}
\hline\noalign{\smallskip}
\multicolumn{1}{c}{No.} &Constraints   \\
\noalign{\smallskip}\hline\noalign{\smallskip}
\label{re:theta1}&$\theta_{1}= 2\delta_\mathtt{I}/(1-\rho)+\delta_\mathtt{p}$\\
\label{re:theta2}&$\theta_{2}= \theta_{1}+\delta_{2}/(1-\rho)+\delta_\mathtt{p}$\\
\label{re:theta3}&$\theta_{3}= \theta_{2}+\delta_{0}$\\
\label{re:theta4}&$\theta_{4}= (\delta_{3}-\delta_{1}+2\delta_\mathtt{I})/(1-\rho)+\delta_\mathtt{p}$\\
\label{re:theta5}&$\theta_{5}= \theta_{4}+\delta_{4}/(1-\rho)+\delta_\mathtt{p}$\\
\label{re:sigma1}&$\sigma_{1}= \delta_{10}/(1-\rho)+\delta_\mathtt{d}$\\
\label{re:sigma2}&$\sigma_{2}= \delta_{15}/(1+\rho)-\delta_\mathtt{d}-\delta_{10}/(1-\rho)$\\
\label{re:sigma2g}&$\sigma_{2}\geqslant (k_\mathtt{pls}-1)(\tau_0+2\delta_{\mathtt{I}})/(1-\rho)$\\
\label{re:sigma3}&$\sigma_{3}= 2\delta_\mathtt{p}+\delta_{11}/(1-\rho)$\\
\label{re:sigma4}&$\sigma_{4}= 2\delta_\mathtt{p}+\delta_{17}/(1-\rho)$\\
\label{re:sigma5}&$\sigma_{5}= \sigma_{7}+\delta_{14}/(1-\rho)+\delta_\mathtt{p}$\\
\label{re:sigma6}&$\sigma_{6}= \sigma_{1}+\sigma_{2}+\sigma_{3}+(\tau_0+\delta_{1}+\delta_\mathtt{I})(1+\rho)+\delta_\mathtt{p}$\\
\label{re:sigma7}&$\sigma_{7}=\delta_{13}/(1-\rho)+\delta_\mathtt{d}$\\
\label{re:sigma8}&$\sigma_{8}=\delta_{11}/(1+\rho)$\\
\label{re:sigma9}&$\sigma_{9}=(\sigma_{3}-\sigma_{8}+\delta_\mathtt{d})(1+\rho)$\\
\label{re:sigma10}&$\sigma_{10}=\delta_{12}(1+\rho)$\\
\label{re:sigma11}&$\sigma_{11}=(\delta_{7}+\sigma_{9}+2\rho\sigma_{10})(1+\rho)+\delta_\mathtt{p}$\\
\label{re:sigma12}&$\sigma_{12}=\sigma_{3}+\sigma_{4}+\sigma_{10}+\sigma_{11}+\delta_{0}$\\
\label{re:sigma13}&$\sigma_{13}=2\delta_\mathtt{I}/(1-\rho)+\delta_\mathtt{p}+\sigma_{6}$\\
\label{re:sigma14}&$\sigma_{14}=\sigma_{6}+(k_\mathtt{pls}-1)T_\mathtt{max}$\\
\label{re:alpha}&$\alpha=(\lfloor (n_0-2f_0-1)/f_0 \rfloor+1)^{-1}$\\
\label{re:epsilonb}&$\epsilon_\mathtt{b}=11\varepsilon_{0}+\rho(3\theta_{1}+2\theta_{5}+4\theta_{4}-4\theta_{3}+T_\mathtt{max})$\\
\label{re:deltaI1}&$\delta_\mathtt{I}\geqslant \max\{\sigma_{11}+2\varepsilon_{0}+2\rho\tau_0,\varepsilon_{2}+2\rho k_\mathtt{pls}T_\mathtt{max}\}$\\
\label{re:Tmin}&$T_\mathtt{min}=(\tau_0-\delta_{6})/(1+\rho)-\delta_\mathtt{p}$\\
\label{re:Tmax}&$T_\mathtt{max}=(\tau_0+\delta_{6})/(1-\rho)+\delta_\mathtt{p}$\\
\label{re:deltac}&$\Delta_\mathtt{C}=\delta_{14}/(1-\rho)+\delta_\mathtt{p}$\\
\label{re:varepsilon1varrho1}&$\varepsilon_{1}> 2\epsilon_\mathtt{b}/(1-\alpha)$, $\varrho_{1}= \rho+\varepsilon_{1}/T_\mathtt{min}$\\
\label{re:Delta1}&$\Delta_{1}=\Delta_\mathtt{C}+3k_\mathtt{pls}\tau_0/\eta_1+\sigma_{14}$\\
\noalign{\smallskip}\hline
\end{tabular}
\end{table}

Besides the constant parameters, each node $i\in U$ also uses some local variables in running the algorithms.
In the analysis, we use $x^{(i)}$ to denote the local variable $x$ used in $i\in U$ when it is needed to differentiate the different nodes.
And the value of $x$ (or $x^{(i)}$) at $t$ is denoted as $x(t)$ (or $x^{(i)}(t)$).
For example, the value of $\textit{offset}_i$ in running the $\mathtt{BFT\_SYNC}$ algorithm in node $i\in U$ at $t$ can be denoted as $\textit{offset}_i^{(i)}(t)$ (or simplified as $\textit{offset}_i(t)$).
Also, we assume that each line of the algorithms is atomically executed.
And when any line of the algorithms is being executed in $i\in U$ at $t$, we assume that $x^{(i)}(t)$ takes the value after the execution of this line.

As is shown in the algorithms, the timers can also be represented as local variables.
In considering stabilization, all the local variables might have arbitrary values at $t_0$.
In the algorithms, as we always check the timers with value ranges as small as possible, all these timers can be locally stabilized within their scheduled ticks.
And for all the other local variables, it is trivial to show that the history values recorded before $t_0$ can be overwritten within the maximal scheduled ticks of the timers.
So for convenience, we assume that all the local variables used in the algorithms are overwritten at least once since $t_0$  at some instant $t_\mathtt{C}>t_0$.
With the provided algorithms, we have $t_\mathtt{C}\in [t_0,t_0+\Delta_\mathtt{C}]$, where $\Delta_\mathtt{C}=\delta_{14}/(1-\rho)+\delta_\mathtt{p}$ is called the local recovery time.
As all the local variables used in the algorithms can be recovered from all the possible incorrect values before $t_\mathtt{C}$, a node in $U$ can be referred to as a \emph{correct} node since $t_\mathtt{C}$ (following \cite{Daliot2006Agreement} and \cite{DolevPulseBoundedDelay2007}).

In the analysis, when the clocks are added or subtracted with small quantities (such as the timeouts, the reading errors, the message delays, the adjustment cycles), as $\tau_{max}$ is assumed to be far greater than such quantities, the default (i.e., would be automatically performed in the computer) $\bmod$ operations can be ignored.
Especially in representing the value range of the clocks, we would use the common operators $+$ and $-$ rather than $\oplus$ and $\ominus$.
For example, the value range $[\tau-\delta,\tau+\delta]$ of a clock would actually be $[\tau-\delta+\tau_{max},\tau_{max})\cup[0,\tau+\delta]$ if $\tau-\delta<0$ and would actually be $[\tau-\delta,\tau_{max})\cup[0,\tau+\delta-\tau_{max}]$ if $\tau+\delta\geqslant \tau_{max}$.
But for simplicity, we would rather take the common representation $[\tau-\delta,\tau+\delta]$.
Also, the default rounding operations on the discrete ticks are ignored in handling the multiplication and division operations (one can add an extra tick in each such operation to derive a sufficiently safe configuration).
All these ignored operations (modular and rounding) can be trivially added when needed.

Firstly, for the strong synchronizer, we give the definition of a synchronization point.
\begin{definition}
\label{def_wellsynchronized}
$t$ is a $\delta$-synchronization point iff $\mathcal{L}$ is initially $\delta$-synchronized at $t$, no pulse is being transmitted or processed in $\mathcal{L}$ at $t$, the timers $\tau_{w}$, $\tau_{w}^*$ are all reset at $t$, and no line nor block of the $\mathtt{BFT\_SYNC}$ or $\mathtt{BFT\_PULSE\_SYNC}$ algorithm is being executed at $t$.
\end{definition}

For example, the vertical dotted lines $t=t_1$, $t=t_3$, $t=t_4$, $t=t_6$, and $t=t_7$ in Fig.~\ref{fig:basicprocess} all correspond to some synchronization points.
But $t_2$ and $t_5$ (in Fig.~\ref{fig:basicprocess}) are not synchronization points, since they are covered in some updating spans of the local clocks.
Similarly, in Fig.~\ref{fig:headerbody}, $t_2$, $t_4$, and $t_6$ can be synchronization points while $t_1$, $t_3$, and $t_5$ cannot be.
Generally, with the synchronization points, the synchronization phases in the synchronized system can be well-separated.
For analysis, here we further define some specific synchronization points between the separated synchronization phases.
\begin{definition}
\label{def_forerunner}
$t$ is a $(\delta,\delta',k)$-synchronization point iff $t$ is a $\delta$-synchronization point and $\exists j\in U_1:C_{j}(t)=k\tau_0+\delta'-\delta$.
\end{definition}

\subsection{The basic synchronizer}

In this subsection, we assume the $\mathtt{BFT\_SYNC}$ algorithm runs alone (i.e., all the other algorithms are ignored here), and the system is in an initial $\delta_{\mathtt{I}}$-synchronized state.
With this, we show that the $\mathtt{BFT\_SYNC}$ algorithm can maintain the synchronized state of the system with the strictly separated synchronization phases shown in Fig.~\ref{fig:basicprocess}.
Especially, we show that the synchronous approximate agreement can be simulated in CCBN with $n_0>3f_0$ and $n_1>2f_1$.
The instants $t_x$ (for $x=1,2,\dots,6$) referenced in Lemma~\ref{lemma_sync_maintain0} correspond to the ones shown in Fig.~\ref{fig:basicprocess}.
As is mentioned, this is mainly for the ease of reading and can be further optimized for shorter synchronization cycles when it is needed.
And for simplicity, we do not redefine the parameters given in Table~\ref{tab:parameter_relations} and Table~\ref{tab:related_relations} in the proofs.
The readers can easily check the relations of the parameters used in the proofs with these tables.
By this, we can also avoid the \emph{magic numbers} and \emph{premature calculations} being scattered in the proofs.

\begin{lemma}
\label{lemma_sync_maintain0}
If there is a $(\delta,\delta_{1},k)$-synchronization point $t_0'$ with some $\delta\leqslant\delta_\mathtt{I}$ and $k\in \mathbb Z^+$, then there is a $(\delta',\delta_{1},k+1)$-synchronization point $t'\in [t_0'+T_\mathtt{min},t_0'+T_\mathtt{max})$ with some $\delta'\leqslant\alpha\delta+\epsilon_\mathtt{b}$ and $\mathcal{L}$ is $(2\delta',\rho)$-synchronized during $[t_0',t']$.
\end{lemma}
\begin{IEEEproof}
As $t_0'$ is a $(\delta,\delta_{1},k)$-synchronization point, there is some $j_0\in U_1$ satisfying $C_{j_0}(t_0')=k\tau_0+\delta_{1}-\delta$ and $\forall j\in U:\mathring{d}(C_{j_0}(t_0'),C_j(t_0'))\leqslant \delta$.
So we have $C_j(t_0')\in[k\tau_0+\delta_{1}-2\delta,k\tau_0+\delta_{1}]$ for all $j\in U$.
As $\tau_{w}^{(j)}(t_0')=\tau_{max}$ and no line of the $\mathtt{BFT\_SYNC}$ algorithm is being executed at $t_0'$, the $\tau_{w}^{(j)}$ timer would remain being closed and $L_j$ and $C_j$ would not be adjusted before some line (of the $\mathtt{BFT\_SYNC}$ algorithm, the same below) being executed in $j$ since $t_0'$.

As $C_i(t_0')\in[k\tau_0+\delta_{1}-2\delta,k\tau_0+\delta_{1}]$ in every node $i\in U_0$, $L_i$ and $C_i$ would not be adjusted during $[t_0',t_3)$ with $t_3=t_0'+\theta_{3}\geqslant t_0'+\theta_{2}+\delta_{0}$ (see Table~\ref{tab:parameter_relations} and Table~\ref{tab:related_relations}, the same below).
During $[t_0',t_3)$, as $C_j(t_0')\in[k\tau_0+\delta_{1}-2\delta,k\tau_0+\delta_{1}]$, every node $j\in U_1$ would read the remote clocks $\hat{C}_{i,j}(t_j')$ and write $L_j$ in executing line~\ref{code:backward_sync_read} at some $t_j'\in [t_0',t_1)$ with $t_{1}=t_0'+\theta_{1}$.
Denoting $\tilde{c}_{i,j}(t)=\hat{C}_{i,j}(t)\ominus(C_j(t)-\delta_{5})$, as $\forall i\in U_0,\forall j\in U_1:\mathring{d}(\hat{C}_{i,j}(t),C_i(t))\leqslant \varepsilon_{0} \land \mathring{d}(C_{i}(t),C_{j}(t))\leqslant \delta_0'=\delta+2\rho\theta_{1}$ for all $t\in [t_0',t_1]$, for every $j\in U_1$ we have $\forall i\in U_0:\tilde{c}_{i,j}(t)\in[\tau_j,\tau_j+\delta_1'] \land \mathring{d}(\tilde{c}_{i,j}(t),\tilde{c}_{i,j}(t_1))\leqslant \delta_2'$ with some $\tau_j\in[\delta_{5}-\delta_1',\delta_{5}]$, $\delta_1'=\delta_0'+ 2\varepsilon_{0}<\delta_{5}$, and $\delta_2'=2\rho\theta_{1}+2\varepsilon_{0}$ for all $t\in [t_0',t_{1}]$.
So with the basic properties of the FTA function, as $n_0>3f_0$ and $\forall j\in U_1:t_j'\in [t_0',t_1)$, we have $|\textit{offset}_j^{(j)})(t_j')|\leqslant \delta_1'$ and $\forall j_1,j_2\in U_1:\mathring{d}(L_{j_1}(t_{1}),L_{j_2}(t_{1}))\leqslant \delta_3'$ with $\delta_3'=\delta_1'/2+2\delta_2'$.
Then, every node $j\in U_1$ would execute line~\ref{code:backward_sync_offset_update} during $[t_{1},t_{2})$ with $t_{2}=t_0'+\theta_{2}$.
So we have $\forall j_1,j_2\in U_1:\mathring{d}(C_{j_1}(t_{2}),C_{j_2}(t_{2}))\leqslant\delta_3'+2\rho\theta_{2}$ and $\forall j\in U_1:\tau_{w}^{(j)}(t_2)=\tau_{max}$.

Then, every node $j\in U_1$ would not adjust $L_j$ nor $C_j$ and would not set $\tau_{w}^{(j)}$ during $[t_2,t_6)$ with $t_6= t_0'+(\tau_0-\delta_{6}) /(1+\rho)$.
So we have $\forall j_1,j_2\in U_1:\mathring{d}(C_{j_1}(t),C_{j_2}(t))\leqslant\delta_3'+2\rho(\tau_0-\delta_{6}) /(1+\rho)$ for all $t\in [t_2,t_6)$.
So as $t_3-t_2\geqslant \delta_{0}$, we have $\forall i\in U_0,\forall j\in U_1:\mathring{d}(\hat{C}_{j,i}(t),C_j(t))\leqslant \varepsilon_{0}$ for all $t\in [t_3,t_6)$.
And every node $i\in U_0$ would read the remote clocks $\hat{C}_{j,i}$ and write $L_i$ with the median of the remote readings from $V_1$ in executing line~\ref{code:sync_read} at some $t_i'\in[t_3,t_4)$ with $t_4=t_0'+\theta_{4}$.
Also, denoting $\tilde{c}_{j,i}(t)=\hat{C}_{j,i}(t)\ominus(C_i(t)-\delta_{6})$, as $\forall i\in U_0,\forall j\in U_1:\mathring{d}(\hat{C}_{j,i}(t),C_j(t))\leqslant \varepsilon_{0} \land \mathring{d}(C_{i}(t),C_{j}(t))\leqslant \delta_4'=\delta_3'+2\rho(t_4-t_1)$ for all $t\in[t_3,t_4]$, for every $i\in U_0$ we have $\forall j\in U_1:\tilde{c}_{j,i}(t)\in[\tau_i,\tau_i+\delta_5'] \land \mathring{d}(\tilde{c}_{j,i}(t),\tilde{c}_{j,i}(t_4))\leqslant \delta_6'$ with some $\tau_i\in [\delta_{6}-\delta_5',\delta_{6}]$, $\delta_5'=\delta_4'+ 2\varepsilon_{0}\leqslant \delta_{6}$, and $\delta_6'=2\rho(t_4-t_3)+2\varepsilon_{0}$ for all $t\in [t_3,t_4]$.
So with the basic properties of the median function, as $n_1>2f_1$ and $\forall i\in U_0:t_i'\in [t_3,t_4)$, we have $\forall i,j\in U:\mathring{d}(L_{i}(t_4),L_{j}(t_4))\leqslant \delta_7'$ with $\delta_7'=\delta_5'+2\delta_6'$.
Then, every node $i\in U_0$ would execute line~\ref{code:sync_offset_update} during $[t_4,t_5)$ with $t_5=t_0'+\theta_{5}\leqslant t_6-\delta_{0}$.
So we have $\forall i_1,i_2\in U:\mathring{d}(C_{i_1}(t_5),C_{i_2}(t_5))\leqslant\delta_8'$ with $\delta_8'=\delta_7'+2\rho(t_5-t_4)$ and $\forall i\in U:\tau_{w}^{(j)}(t_5)=\tau_{max}$.

Then, as $t_6-t_5\geqslant \delta_{0}$, we have $\forall i\in U_0,j\in U_1:\mathring{d}(\hat{C}_{i,j}(t),C_i(t))\leqslant \varepsilon_{0} \land \mathring{d}(\hat{C}_{j,i}(t),C_j(t))\leqslant \varepsilon_{0}$ for all $t\in [t_6,t_7]$ with $t_7\geqslant t_6$ being the earliest instant satisfying $C_j(t_7)=(k+1)\tau_0+\delta_{1}$ for some $j\in U_1$.
So there is a $\delta'$-synchronization point $t'\in [t_6,t_7]$ satisfying $\exists j_0'\in U_1:C_{j_0'}(t')=(k+1)\tau_0+\delta_{1}-\delta'$.
As the maximal difference of the logical clocks of the nodes in $U$ is within $2\delta'$ during $[t_0',t_7]$, in which every node in $U$ adjusts its logical clock at most once with no more than $2\delta'\leqslant \delta_{6}$ clock-adjustment, $\mathcal{L}$ is $(2\delta',\rho)$-synchronized during $[t_0',t']$ with $t'\in[t_0'+T_\mathtt{min},t_0'+T_\mathtt{max})$.
\end{IEEEproof}

\begin{corollary}
\label{corollary_remainsync}
If the premise of Lemma~\ref{lemma_sync_maintain0} holds, $\mathcal{L}$ would be $(2\delta_{(c)},\rho_{(c)})$-synchronized since $t+cT_\mathtt{max}$ with $\rho_{(c)}\leqslant \rho+2\delta_{(c)}/T_\mathtt{min}$ and $\delta_{(c)}\leqslant\alpha^c \delta+\epsilon_\mathtt{b}/(1-\alpha)$.
\end{corollary}
\begin{IEEEproof}
Denote $\delta_{(0)}=\delta$ and $\delta_{(1)}=\delta'$ for the parameters $\delta$ and $\delta'$ used in Lemma~\ref{lemma_sync_maintain0}, respectively.
By applying Lemma~\ref{lemma_sync_maintain0}, we have $\delta_{(1)}=\alpha \delta_{(0)}+\epsilon_\mathtt{b}$ with $t_{(1)}\in [t+T_\mathtt{min},t+T_\mathtt{max})$.
As the premise of Lemma~\ref{lemma_sync_maintain0} also holds for $t_{(1)}$, we have $\delta_{(2)}=\alpha \delta_{(1)}+\epsilon_\mathtt{b}$ with $t_{(2)}\in [t_{(1)}+T_\mathtt{min},t_{(1)}+T_\mathtt{max})$.
Iteratively, we have $\delta_{(c)}=\alpha^c \delta+\epsilon_\mathtt{b}(1-\alpha^c)/(1-\alpha)\leqslant\alpha^c \delta+\epsilon_\mathtt{b}/(1-\alpha)$ with $t_{(c)}\in [t+cT_\mathtt{min},t+cT_\mathtt{max})$.
As $\delta_{(0)}\leqslant \delta_\mathtt{I}$, we have $\delta_{(c')}\leqslant \delta_{(c'-1)}$ for all $c'\in \mathbb Z^+$.
So the synchronization precision $2\delta_{(c)}$ and the accuracy $\rho_{(c)}\leqslant \rho+2\delta_{(c)}/T_\mathtt{min}$ can be maintained in $\mathcal{L}$ since $t_{(c)}$.
\end{IEEEproof}

Notice that in the provided algorithms, we use the basic FTA functions in simulating the basic approximate agreement, which achieves the basic convergence rate $\alpha_0=1/2$.
For faster convergence, the FTA functions can be replaced as the advanced ones to achieve the convergence rate $\alpha=(\lfloor (n_0-2f_0-1)/f_0 \rfloor+1)^{-1}$ (see \cite{Dolev1986Approximate} for details).
For example, with $n_0>5f_0$, we would get a better convergence rate $\alpha\leqslant 1/4$.
And this is in line with our basic system settings.
\subsection{The strong synchronizer and strong detector}

Now we show that with the strong synchronizer and the strong detector, if a node $j\in U_1$ does not detect the undesired system state, i.e., $\textit{alerted}_j^{(j)}(t)=0$ holds for some $t$, then $\mathcal{L}$ can be deterministically synchronized in a finite time.
In this subsection, we assume that only the $\mathtt{BFT\_SYNC}$, $\mathtt{BFT\_PULSE\_SYNC}$, and $\mathtt{S\_DETECTOR}$ algorithms run.

Firstly, denoting the always guarded condition in line~\ref{code:pulse_sync_preempitve_con} of the $\mathtt{BFT\_PULSE\_SYNC}$ algorithm as $\mathtt{A_1}$ and the lines from \ref{code:pulse_sync_preempitve_start} to \ref{code:pulse_sync_preemptive_stop} of the same algorithm in responding to the satisfaction of $\mathtt{A_1}$ as $\mathtt{B_1}$, we give the definition of the semi-synchronous execution of the $\mathtt{B_1}$ block.

\begin{definition}
\label{def_synchronous}
The nodes in $U_1$ perform a $\delta$-synchronous execution of the $\mathtt{B_1}$ block during $[t,t']$ iff for every node $j\in U_1$, there is an execution of the $\mathtt{B_1}$ block during $[t,t']$ such that for every line $l\in \mathtt{B_1}$, $l$ is not preempted or canceled and the execution instants of $l$ are at most $\delta$ apart in all nodes of $U_1$.
\end{definition}

Analogously, denoting the always guarded condition in line~\ref{code:pulse_sync_preempitve_con0} of the $\mathtt{BFT\_PULSE\_SYNC}$ algorithm as $\mathtt{A_0}$ and the lines from \ref{code:pulse_sync_preempitve_start0} to \ref{code:pulse_sync_preemptive_stop0} as $\mathtt{B_0}$, we can also define the semi-synchronous execution of the $\mathtt{B_0}$ block by taking place $U_1$ as $U_0$.
Now we first show that the $\mathtt{A_1}$ condition would not be satisfied very frequently, and thus the $\mathtt{B_1}$ block would be eventually executed in the semi-synchronous way when the $\mathtt{A_1}$ condition is satisfied in all nodes of $U_1$ during a sufficiently short period.

\begin{lemma}
\label{lemma_sync_separated}
If the $\mathtt{A_1}$ condition is satisfied in nodes $j_1,j_2\in U_1$ ($j_1$ and $j_2$ can be the same or different nodes) at $t_{1}$ and $t_{2}$ with $t_{2}>t_{1}$, then $t_{2}-t_{1}\notin(\sigma_{1},\sigma_{2}]$.
\end{lemma}
\begin{IEEEproof}
Denote the sets $P_{k'}$ satisfying the $\mathtt{A_1}$ condition at $t_{1}$ and $t_{2}$ as $P_{k_1}$ and $P_{k_2}$, respectively.
As $|P_{k'}|\geqslant n_0-2f_0$, there are at least $n_0-3f_0$ nonfaulty nodes in every such $P_{k'}$.
As $n_0>5f_0$, there exists $i_0\in U_0\cap P_{k_1}\cap P_{k_2}$ for otherwise it can only be $2(n_0-3f_0)\leqslant n_0-f_0$.
For every such $i_0$, if its pulse is received in any $j_1\in U_1$ at some $t''$, this pulse can only be sent at some $t'\in[t''-\delta_\mathtt{d},t'']$ and thus can only be received in $j_2\in U_1$ during $[t',t'+\delta_\mathtt{d}]$.
So if the $\mathtt{A_1}$ condition is satisfied in $j_1$ and $j_2$ with receiving the same pulse of $i_0$, then $t_{2}-t_{1}\leqslant \sigma_{1}$ holds.
Otherwise, if two pulses are sent by any $i\in U_0$ at $t_{1}'$ and $t_{2}'$ with $t_{2}'>t_{1}'$, with the condition in line~\ref{code:pulse_sync_pulse_con}, we have $t_{2}'-t_{1}'\geqslant \epsilon$ with $\epsilon=\delta_{15}/(1+\rho)$.
So within any $\epsilon-\delta_\mathtt{d}$ time, at most one pulse from $i_0$ would be received in the nodes of $U_1$.
So if $t_{2}-t_{1}> \delta_\mathtt{d}+\delta_{10}/(1-\rho)$, we have $t_{2}-t_{1}> \epsilon-\delta_\mathtt{d}-\delta_{10}/(1-\rho)$.
\end{IEEEproof}

\begin{lemma}
\label{lemma_sync_synchronous}
If the $\mathtt{A_1}$ condition is satisfied in every $j\in U_1$ at some $t_j'\in[t_{0}',t_{0}'+\sigma_{2}]$, then there exists $t^*\in [t_{0}',t_{0}'+\sigma_{1}+\sigma_{2}]$ such that the $\mathtt{A_1}$ condition is satisfied in every $j\in U_1$ at some $t_j^*\in[t^*-\sigma_{1},t^*]$ and all nodes in $U_1$ would perform a $\delta$-synchronous execution of the $\mathtt{B_1}$ block during $[t^*-\sigma_{1},t^*+\sigma_{3}]$ with $\delta=\sigma_{1}+2\rho\sigma_{3}$.
\end{lemma}
\begin{IEEEproof}
As the $\mathtt{A_1}$ condition is satisfied in every $j\in U_1$ at some $t_j'\in[t_{0}',t_{0}'+\sigma_{2}]$, with Lemma~\ref{lemma_sync_separated} we have $t_{j}^*-t_j'\in[0,\sigma_{1}]$ with $t_{j}^*$ being the last instant when the $\mathtt{A_1}$ condition is satisfied in $j$ before $t_{0}'+\sigma_{2}$.
Again with Lemma~\ref{lemma_sync_separated}, we have $\max_{j_1,j_2\in U_1}|t_{j_1}'-t_{j_2}'|\leqslant \sigma_{1}$.
So we have $\max_{j_1,j_2\in U_1}|t_{j_1}^*-t_{j_2}^*|\leqslant 2\sigma_{1}$.
As $\sigma_{2}\geqslant 2\sigma_{1}$, we have $\max_{j_1,j_2\in U_1}|t_{j_1}^*-t_{j_2}^*|\leqslant \sigma_{1}$ also with Lemma~\ref{lemma_sync_separated}.
Now as the $\mathtt{A_1}$ condition would not be satisfied in every node $j$ during $(t_{j}^*,t_{j}^*+\sigma_{2}]$, all nodes in $U_1$ would perform a $\delta$-synchronous execution of the $\mathtt{B_1}$ block during $[t^*-\sigma_{1},t^*+\sigma_{3}]$ with some $t^*\in[t_{0}',t_{0}'+\sigma_{1}+\sigma_{2}]$.
\end{IEEEproof}

Now we show that the semi-synchronous approximate agreement can be initiated if the strong detector cannot detect the undesired system state in any node $j\in U_1$.
For the ease of reading, like Lemma~\ref{lemma_sync_maintain0}, here the instants $t_x$ (for $x=1,2,\dots,6$) referenced in Lemma~\ref{lemma_sync_converge0} correspond to the ones shown in Fig.~\ref{fig:headerbody}.

\begin{lemma}
\label{lemma_sync_converge0}
If there is $j_0\in U_1$ with $\textit{alerted}_j^{(j_0)}(t)=0$ at some $t\geqslant t_{0}+\sigma_{5}$, then there is a $(\delta,\delta_{1},kk_\mathtt{pls}+1)$-synchronization point in $[t-\sigma_{5},t+\sigma_{6}]$ with some $\delta\leqslant\delta_\mathtt{I}$ and $k\in \mathbb Z^+$.
\end{lemma}
\begin{IEEEproof}
Firstly, as $\textit{alerted}_j^{(j_0)}(t)=0$, the condition of line~\ref{code:detector_satis} of the $\mathtt{S\_DETECTOR}$ algorithm is satisfied at some $t'\in[t-\delta_{14}/(1-\rho)-\delta_\mathtt{p},t]$ in $j_0$.
As $t'\geqslant t_{0}+\sigma_{7}$ and $|P_k^{(j_0)}(t')|\geqslant n_0-f_0$ hold for some $k$, at least $n_0-2f_0$ nodes in $U_0$ send their pulses with their local clocks being $\tau=kk_\mathtt{pls}\tau_0$ during $[t'-\sigma_{7},t']$ with $t'-\sigma_{7}\geqslant t_{0}$.
Denoting $P=P_k^{(j_0)}(t')\cap U_0$, we have $|P|\geqslant n_0-2f_0$ and $P$ being a pulsing clique in $U_0$.
So for every node $j\in U_1$ we have $P\subseteq P_{k}^{(j)}(t_j')$ with some $t_j'\in[t_{0}',t_{0}'+\sigma_{7}+\delta_\mathtt{d}]$ and some $t_{0}'\in [t'-\sigma_{7},t']$.
So as $(\sigma_{7}+\delta_\mathtt{d})(1+\rho)\leqslant\delta_{10}$, the $\mathtt{A_1}$ condition would be satisfied at $t_j'$ for every node $j\in U_1$ with the same $k$.

Then, by applying Lemma~\ref{lemma_sync_separated} and Lemma~\ref{lemma_sync_synchronous}, as the $\mathtt{A_1}$ condition is satisfied in every $j\in U_1$ at $t_j'\in[t_{0}',t_{0}'+\sigma_{7}+\delta_\mathtt{d}]$, the $\mathtt{B_1}$ block would be semi-synchronously executed in every node $j\in U_1$ during $[t_j^*,t_j^*+\sigma_{3}]$.
In executing these lines in every node $j\in U_1$, as only the values in $[0,\delta_{7}]$ would be input to ${\tau_{i,j}'}^{(j)}$ in executing line~\ref{code:pulse_sync_input} (of the $\mathtt{BFT\_PULSE\_SYNC}$ algorithm, the same below), $C_j(t_j'')\in[{\tau'}^{(j)}(t_j''),{\tau'}^{(j)}(t_j'')+\delta_{7}]$ trivially holds when $C_j$ is adjusted by executing line~\ref{code:pulse_sync_preemptive_do} at some $t_j''\in[t_j^*+\sigma_{8},t_j^*+\sigma_{3}]$.
As $\forall j\in U_1:k^{*(j)}(t_j'')=k$, we have $\forall j_1,j_2\in U_1:{\tau'}^{(j_1)}(t_{j_1}'')={\tau'}^{(j_2)}(t_{j_2}'')=kk_\mathtt{pls}\tau_0+\delta_{8}$.
So with Lemma~\ref{lemma_sync_synchronous}, as $t_j''\in [t_j^*+\sigma_{8},t_j^*+\sigma_{3}]$ with some $t_j^*\in[t^*-\sigma_{1},t^*]$ with some $t^*\in [t_{0}',t_{0}'+\sigma_{1}+\sigma_{2}]$, we have $\forall j\in U_1:C_{j}(t_j'')-(kk_\mathtt{pls}\tau_0+\delta_{8})\in[0,\delta_{7}]$ and $\forall j_1,j_2\in U_1:\mathring{d}(C_{j_1}(t),C_{j_2}(t))\leqslant \delta_1'$ for all $t\in [t^*,t_1]$ with $\delta_1'=\delta_{7}+\sigma_{9}$ and $t_1=t^*+\sigma_{3}$.

So, $\forall j_1,j_2\in U_1:\mathring{d}(C_{j_1}(t),C_{j_2}(t))\leqslant \delta_2'$ holds for all $t\in[t_1,t_2]$ with $\delta_2'=\delta_1'+2\rho\sigma_{10}$ and $t_2\in[t_1, t_1+\sigma_{10}]$ being the earliest instant satisfying $C_{j}(t_2)=kk_\mathtt{pls}\tau_0+\delta_{12}$ for some $j\in U_1$.
In other words, all nodes in $U_1$ would have been coarsely synchronized by the pulsing clique $P$ with a precision no worse than $\delta_2'$ at the statically scheduled pulsing instants.
As all attempts to write $L_j$ or $C_j$ in the $\mathtt{BFT\_SYNC}$ algorithm would be cancelled before $C_j$ reaching $(kk_\mathtt{pls}+1)\tau_0$, all the nodes in $U_1$ would send their pulses during $[t_2,t_3]$ with $t_3= t_2+\sigma_{11}$.

Then, similar to the proof of Lemma~\ref{lemma_sync_synchronous}, the $\mathtt{A_0}$ condition would be satisfied at some $t_i^*\in [t_2,t_4]$ in every node $i\in U_0$ and the $\mathtt{B_0}$ block would be semi-synchronously executed in $U_0$ during $[t_2,t_5]$ with $t_4=t_3+\delta_{0}$ and $t_5=t_4+\sigma_{4}$.
Thus, every node $i\in U_0$ would remotely read the synchronized local clocks of $U_1$ and set $C_i$ with these readings during $[t_2,t_5]$.
And with line~\ref{code:pulse_sync_preemptive_stop0}, all attempts to write $L_i$ or $C_i$ in the $\mathtt{BFT\_SYNC}$ algorithm would be cancelled before $C_i$ reaching $(kk_\mathtt{pls}+1)\tau_0$.

So, we have $\forall i,j\in U:\mathring{d}(C_{i}(t),C_{j}(t))\leqslant \delta=\delta_2'+2\varepsilon_{0}+2\rho\tau_0$ for all $t\in[t_5,t_6]$, where $t_6$ is the first instant that some node $j\in U_1$ satisfying $C_{j}(t)=(kk_\mathtt{pls}+1)\tau_0+\delta_{1}-\delta$ since $t_2$.
As every $C_j$ is updated as some value no more than $kk_\mathtt{pls}\tau_0+\delta_{12}$ at some $t\in[t^*,t_6]$, we have $t_6-t^*\geqslant (\tau_0-\delta_{12}+\delta_{1}-\delta)/(1+\rho)$.
So with $t_5=t_4+\sigma_{4}=t_3+\delta_{0}+\sigma_{4}=t_2+\sigma_{11}+\delta_{0}+\sigma_{4}\leqslant t_1+\sigma_{10}+\sigma_{11}+\delta_{0}+\sigma_{4}=t^*+\sigma_{12}$, we have $t_6\geqslant t_5+\delta_{0}$ and thus $t_6$ is a $\delta$-synchronization point satisfying $t_6\in[t-\sigma_{5},t+\sigma_{6}]$.
\end{IEEEproof}

Then, it is easy to see that the semi-synchronous approximate agreement can bring the system to a desired synchronized state at the beginning of the new synchronization header.
\begin{lemma}
\label{lemma_sync_converge1}
If there is a $(\delta,\delta_{1},kk_\mathtt{pls}+1)$-synchronization point $t_0'\geqslant t_0$ with any $\delta\in[\varepsilon_{1}/2,\delta_\mathtt{I}]$ and $k\in \mathbb Z^+$, then there is a $(\delta',0,(k+1)k_\mathtt{pls})$-synchronization point $t_0''\in [t_0'+cT_\mathtt{min}-\delta_1/(1-\rho),t_0'+cT_\mathtt{max})$ with $\delta'\leqslant\varepsilon_{1}/2$, $c=k_\mathtt{pls}-1$, $\mathcal{L}$ is $(2\delta,\rho+2\delta/T_\mathtt{min})$-synchronized during $[t_0',t_0'']$, and every node in $U_0$ sends a pulse during $[t_0'',t_0''+2\delta'/(1-\rho)+\delta_\mathtt{p}]$.
\end{lemma}
\begin{IEEEproof}
As $t_0'$ is a $(\delta,\delta_{1},kk_\mathtt{pls}+1)$-synchronization point, no line of the $\mathtt{BFT\_PULSE\_SYNC}$ algorithm would be executed during $[t_0',t_1]$, where $t_1\geqslant t_0'$ is the earliest instant satisfying $C_{i}(t_1)=kk_\mathtt{pls}\tau_0+k_\mathtt{pls}\tau_0$ for some $i\in U$.
So, with the proof of Corollary~\ref{corollary_remainsync}, there is a $\delta'$-synchronization point $t_0''\in [t_1-\delta',t_1]$ with $\delta'\leqslant\alpha^{c} \delta+\epsilon_\mathtt{b}/(1-\alpha)$, $\exists j_0\in U_1:C_{j_0}(t_0'')=(k+1)k_\mathtt{pls}\tau_0-\delta'$, and $\mathcal{L}$ is $(2\delta,\rho+2\delta/T_\mathtt{min})$-synchronized during $[t_0',t_0'']$.
So we have $\delta'\leqslant \varepsilon_{1}/2$.
And with such a $(\delta',0,(k+1)k_\mathtt{pls})$-synchronization point $t_0''$, the condition in line~\ref{code:pulse_sync_pulse_con} of the $\mathtt{BFT\_PULSE\_SYNC}$ algorithm would be satisfied in every node $i\in U_0$ during $[t_0'',t_0''+2\delta'/(1-\rho)]$.
So every node $i\in U_0$ would send a pulse during $[t_0'',t_0''+2\delta'/(1-\rho)+\delta_\mathtt{p}]$.
\end{IEEEproof}

Then, if the synchronization header in the following synchronization cycle can work as good as the basic synchronization round, with the proof of Corollary~\ref{corollary_remainsync}, $\mathcal{L}$ would be $(\varepsilon_{1},\varrho_{1})$-synchronized since $t+\delta'$.
Now we show that the synchronization header is as good as a basic synchronization round in maintaining the synchronized state of a $(\varepsilon_{1},\varrho_{1})$-synchronized system.

\begin{lemma}
\label{lemma_sync_converge2}
If there is a $(\delta,0,kk_\mathtt{pls})$-synchronization point $t_0'\geqslant t_0+2T_\mathtt{max}$ with $\delta=\varepsilon_{1}/2$, $k\in \mathbb Z^+$, and every node in $U_0$ sends a pulse during $[t_0',t_0'+2\delta/(1-\rho)+\delta_\mathtt{p}]$, then there is a $(\delta,\delta_{1},kk_\mathtt{pls}+1)$-synchronization point $t_0''\in [t_0'+T_\mathtt{min}+\delta_1/(1+\rho),t_0'+T_\mathtt{max}+\delta_1/(1-\rho)]$ and $\mathcal{L}$ is $(\varepsilon_{1},\rho)$-synchronized during $[t_0',t_0'']$.
\end{lemma}
\begin{IEEEproof}
As every node in $U_0$ sends a pulse during $[t_0',t_0'+2\delta/(1-\rho)+\delta_\mathtt{p}]$, the pulses of all nodes in $U_0$ can all be received in every $j\in U_1$ during $[t_0',t_0'+2\delta/(1-\rho)+\delta_\mathtt{d}]$.
Thus, all nodes in $U_1$ would satisfy the $\mathtt{A_1}$ condition and semi-synchronously execute the $\mathtt{B_1}$ block during $[t_0',t_0'+\sigma_{13}]$, just like the ones shown in the proof of Lemma~\ref{lemma_sync_converge0}.
Thus, with the sufficiently large $\delta_{7}$, a round of synchronous approximate agreement is simulated during $[t_0',t_0'+\sigma_{13}]$.
And with the sufficiently large $\delta_{1}$ (just for clearness), as the $\mathtt{BFT\_SYNC}$ algorithm cannot adjust the clocks of every $j\in U_1$ before $t_0'+2\delta/(1-\rho)+\delta_\mathtt{d}$ in this round, only the $\mathtt{BFT\_PULSE\_SYNC}$ algorithm works in every $j\in U_1$ during this round.
As $\forall i_1,i_2\in U_1:\mathring{d}(C_{i_1}(t),C_{i_2}(t))\leqslant \varepsilon_{1}$ for all $t\in[t_0',t_0'+2\delta/(1-\rho)+\delta_\mathtt{p}+\sigma_{6}]$, we have ${\tau}^{(j)}(t_j')\in [0,\delta_{7}]$ when the line~\ref{code:pulse_sync_raw_input} is executed at some $t_j'\in[t_0',t_0'+\sigma_{13}]$ in every $j\in U_1$.
So, similar to the proof of Lemma~\ref{lemma_sync_maintain0}, we still have $\forall i,j\in U:\mathring{d}(C_{i}(t_0''),C_{j}(t_0''))\leqslant\alpha\delta+\epsilon_\mathtt{b}\leqslant \delta$ when some $j_0\in U_1$ satisfying $C_{j_0}(t_0'')=(k+1)\tau_0+\delta_{1}-\delta$.
\end{IEEEproof}

\begin{theorem}
\label{theorem_sync0}
If there is $j_0\in U_1$ with $\textit{alerted}_j^{(j_0)}(t)=0$ at any $t\geqslant t_{0}+\sigma_{5}$, then $\mathcal{L}$ would be $(\varepsilon_{1},\varrho_{1})$-synchronized since some $t'\in[t,t+\sigma_{14}]$.
\end{theorem}
\begin{IEEEproof}
As $\textit{alerted}_j^{(j_0)}(t)=0$, by applying Lemma~\ref{lemma_sync_converge0}, there is a $(\delta_\mathtt{I},\delta_{1},kk_\mathtt{pls}+1)$-synchronization point $t_0'\in[t-\sigma_{5},t+\sigma_{6}]$ with some $k\in \mathbb Z^+$.
So with Lemma~\ref{lemma_sync_converge1} and Lemma~\ref{lemma_sync_converge2}, there is a $(\varepsilon_{1}/2,\delta_{1},kk_\mathtt{pls}+1)$-synchronization point $t_0'''\in [t_0''+T_\mathtt{min}+\delta_1/(1+\rho),t_0''+T_\mathtt{max}+\delta_1/(1-\rho)]$ with $t_0''\in [t_0'+cT_\mathtt{min}-\delta_1/(1-\rho),t_0'+cT_\mathtt{max})$ and $c=k_\mathtt{pls}-1$, and $\mathcal{L}$ is $(\varepsilon_{1},\rho+\varepsilon_{1}/T_\mathtt{min})$-synchronized during $[t_0'',t_0''']$.
Then, by iteratively applying Lemma~\ref{lemma_sync_converge1} and Lemma~\ref{lemma_sync_converge2}, the conclusion is satisfied with $t'=t_0''$.
\end{IEEEproof}

\subsection{The basic corrector}

As there might be no synchronization point nor any initially $\delta_\mathtt{I}$-synchronized state, some kind of corrector is employed.
As is introduced, the basic corrector comprises a clock merger and the alien clocks.
As the alien clocks are assumed to be synchronized when $\mathcal{L}$ is not synchronized, we mainly study the basic clock merger.
Like \cite{DolevWelchSelf2004}, here we always assume $|U_1|=n_1-f_1$.
Namely, when the number of the actually faulty nodes in $V_1$ is less than $f_1$, some nonfaulty nodes in $V_1$ can be viewed as the faulty ones.
Upon this, when all nodes in $U_1$ are synchronized, as all actually nonfaulty nodes in $V_1$ can be synchronized by the pulsing cliques, the overall stabilization of the system is trivial.

For convenience, we assume the alien clocks $Y_j$ for all $j\in U_1$ satisfying $|Y_j(t)-t|\leqslant \varepsilon_2/2\leqslant e_0$ when $\mathcal{L}$ is not stabilized.
This kind of $Y_j$ clocks are easy to be realized.
For example, we can simply realize $Y_j$ with the remote readings of $R_{z,\mathtt{s}(j)}$ in every node $j\in U_1$.
Concretely, each manager node $s\in S$ can be configured as a synchronization client with some WAN node $z\in Z$ being configured as the synchronization server.
Here, $z\in Z$ can be an external synchronization station (or a multi-source time server or a set of such servers with running a BFT algorithm like $\mathtt{BFT\_READ}$) being connected to $s$ with the minimized safe interface.

Now we show that with some probability $\mathcal{L}$ would be coarsely synchronized and then be finely synchronized when $\mathcal{L}$ is not stabilized.

\begin{lemma}
\label{lemma_corrector_period}
During $[t_1,t_2]$ with $t_1\bmod k_\mathtt{pls}\tau_0= k_\mathtt{pls}\tau_0/2$ and $t_\mathtt{C}+\delta_\mathtt{d}\leqslant t_1\leqslant t_2-3k_\mathtt{pls}\tau_0$, if $\forall t\in[t_1,t_2],\forall j\in U_1: \textit{alerted}_j(t)=1$, then with a probability $\eta_1=2^{3(f_1-n_1)+1}$ that $\mathcal{L}$ would be $(\varepsilon_{1},\varrho_{1})$-synchronized since some $t'\in[t_1,t_2]$.
\end{lemma}
\begin{IEEEproof}
For every node $j\in U_1$, if $\forall t\in [t_1,t_1+k_\mathtt{pls}T_\mathtt{max}]: \textit{pulsed}_j=0$ holds, as the basic-adjustments of $C_j$ are restricted in executing the $\mathtt{BFT\_SYNC}$ algorithm, $C_j(t_{j}')\bmod k_\mathtt{pls}\tau_0= \tau_0$ would be satisfied in $j$ with some $t_{j}'\in[t_1,t_1+k_\mathtt{pls}T_\mathtt{max}]$.
Otherwise, if $\forall t\in [t_1,t_1+k_\mathtt{pls}T_\mathtt{max}]: \textit{pulsed}_j=0$ does not hold, as the execution of the $\mathtt{BFT\_PULSE\_SYNC}$ algorithm has the highest priority, $C_j(t_{j}')\bmod k_\mathtt{pls}\tau_0= \tau_0$ would also be satisfied in $j$ with some $t_{j}'\in[t_1,t_1+(k_\mathtt{pls}+1)T_\mathtt{max}]$.
In both cases, the lines~\ref{code:corrector_coin_toss} and \ref{code:corrector_eor} of the $\mathtt{H\_CORRECTOR}$ algorithm would be executed during $[t_{j}',t_{j}'+\delta_\mathtt{d}]$.
Now as $\forall t\in[t_1,t_2]: \textit{alerted}_j(t)=1$ holds, with at least a probability $1/2$ that $j$ would observe $\textit{EoR}^{(j)}(t)=1$ when executing line \ref{code:corrector_eor} (of the $\mathtt{H\_CORRECTOR}$ algorithm, the same below) during $[t_{j}',t_{j}'+\delta_\mathtt{d}]$.
In this case, the lines~\ref{code:corrector_correction_begin} to \ref{code:corrector_correction_end} would be executed in every $j\in U_1$ during $[t_1,t_1+(k_\mathtt{pls}+1)T_\mathtt{max}+\delta_\mathtt{d}]$.
When the line \ref{code:corrector_correction_end} is executed in $j$, $C_j$ would be written as $Y_j$.
So during $[t_1,t_1+(k_\mathtt{pls}+1)T_\mathtt{max}+\delta_\mathtt{d}]$, line \ref{code:corrector_eor} would be executed in every such $j$ at most twice.
Thus, there is at least a probability $2^{2(f_1-n_1)+1}$ that the local clocks $C_j$ for all $j\in U_1$ are written with $Y_j$ by some $t''\in [t_1,t_1+(k_\mathtt{pls}+1)T_\mathtt{max}+\delta_\mathtt{d}]$.
And with line~\ref{code:corrector_correction_begin}, the basic-adjustments of $C_j$ in every node $j$ would be cancelled since $C_j$ have been written with $Y_j$.
So, during the next execution of the line \ref{code:corrector_eor} in every node $j\in U_1$, as the clock drifts of $C_j$ since $t''$ would be no more than $\rho k_\mathtt{pls}T_\mathtt{max}$, every node $j\in U_1$ would observe $\textit{pulsed}_j=1$ (this step can be omitted if the simplified EOR condition is employed).
Thus, there is at least a probability $1/2$ that $\textit{EoR}^{(j)}(t)=0$ would be observed in $j$ during this execution.
And during this execution, the $C$ clocks of all nodes in $U_1$ would be coarsely synchronized with a precision no worse than $\varepsilon_{2}+2\rho k_\mathtt{pls}T_\mathtt{max}\leqslant \delta_\mathtt{I}$.
In this case, by applying Lemma~\ref{lemma_sync_maintain0}, Corollary~\ref{corollary_remainsync}, Lemma~\ref{lemma_sync_converge1}, and Lemma~\ref{lemma_sync_converge2}, $\mathcal{L}$ would be $(\varepsilon_{1},\varrho_{1})$-synchronized at some $t'\in[t_1,t_2]$.
And as every node $j\in U_1$ would set $\textit{alerted}_j(t')=0$, the EoR condition would not be satisfied since $t'$.
So $\mathcal{L}$ would be $(\varepsilon_{1},\varrho_{1})$-synchronized since $t'$.
\end{IEEEproof}

\begin{lemma}
\label{lemma_corrector_correct}
If some $j_0\in U_1$ satisfies $\textit{alerted}_{j_0}(t)=0$ with some $t\geqslant t_\mathtt{C}+\sigma_{5}$, then with a probability $\eta_2=2^{f_1-n_1+1}$ $\mathcal{L}$ would be $(\varepsilon_{1},\varrho_{1})$-synchronized since some $t'\in[t,t+\sigma_{14}]$.
\end{lemma}
\begin{IEEEproof}
As the the $\mathtt{H\_CORRECTOR}$ algorithm would not be effective (i.e., to execute the lines~\ref{code:corrector_correction_begin} to \ref{code:corrector_correction_end}) when $\textit{alerted}_j=0$ and would not be effective with at least a probability $1/2$ when $\textit{plused}_j=1$ , the result of Theorem~\ref{theorem_sync0} still hold with at least the probability $\eta_2$.
\end{IEEEproof}

\begin{theorem}
\label{theorem_expected}
The expected stabilization time of $\mathcal{L}$ is no more than $\Delta_{1}$.
\end{theorem}
\begin{IEEEproof}
Denote $t_{(0)}$ as the first instant satisfying $t_{(0)}\geqslant t_\mathtt{C}+\delta_\mathtt{d}$ and $t_{(0)}\bmod k_\mathtt{pls}\tau_0= k_\mathtt{pls}\tau_0/2$.
Denote $t_{(k+1)}=t_{(k)}+3k_\mathtt{pls}\tau_0$.
For every $I_k=(t_{(k)},t_{(k)}+3k_\mathtt{pls}\tau_0]$, by applying Lemma~\ref{lemma_corrector_period} and Lemma~\ref{lemma_corrector_correct}, there is at least a probability $\min\{\eta_2,\eta_1\}$ that $\mathcal{L}$ would be $(\varepsilon_{1},\varrho_{1})$-synchronized since some $t'\in I_k$.
So the expected stabilization time of $\mathcal{L}$ is no more than $\Delta_\mathtt{C}+3k_\mathtt{pls}\tau_0/\eta_1+\sigma_{14}$.
\end{IEEEproof}

\subsection{Theoretical results of some concrete instances}
Given the basic system parameters, some concrete configurations of the algorithm parameters that can meet all the constraints (listed in Table~\ref{tab:parameter_relations} and Table~\ref{tab:related_relations}) are shown in Table~\ref{tab:parameters}.
Each column of the values in Table~\ref{tab:parameters} corresponds to some concrete system settings.
For convenience, all the time parameters shown in Table~\ref{tab:parameters} are represented in seconds.
For example, if the value of the time parameter $\tau_0$ is represented as $2.469858$ and the nominal ticking cycle of the hardware clock is $8~ns$, $\tau_0$ should be configured as $\lceil 2.469858\times  125000000\rceil$ ticks.

\begin{table}[htbp]
\caption{A configuration of the constant parameters for the algorithms}
\label{tab:parameters}
% For LaTeX tables use
\begin{tabular}{lllll}
\hline\noalign{\smallskip}
Para. &Case I &Case II &Case III &Case IV   \\
$(n_{0},f_{0})$  &$(6,1)		$&$(100,3)		$&$(6,1) 		$&$(100,3)		$\\
$(n_{1},f_{1})$  &$(3,1)		$&$(3,1)		$&$(5,2) 		$&$(3,1)		$\\
$\rho$                &$0.0001	$&$0.0001	$&$1e-06	$&$1e-06	$\\
$\varepsilon_{0}$   	&$1e-06	$&$1e-06	$&$1e-07	$&$1e-07	$\\
$\varepsilon_{2}$   	&$0.05	$&$0.05	$&$0.001	$&$0.001	$\\
$\delta_\mathtt{p}$ 	&$0.0001	$&$0.0001	$&$2e-05	$&$2e-05	$\\
$\delta_\mathtt{d}$ 	&$0.001	$&$0.001	$&$0.0001	$&$0.0001	$\\
$\delta_{0}$          &$   1	$&$   1	$&$5e-05	$&$5e-05	$\\
$\delta_{1}$        &$  0.105157	$&$  0.104142	$&$  0.002120	$&$  0.002120	$\\
$\delta_{2}$        &$  0.104157	$&$  0.103142	$&$  0.002020	$&$  0.002020	$\\
$\delta_{3}$        &$  1.313691	$&$  1.310645	$&$  0.006231	$&$  0.006230	$\\
$\delta_{4}$        &$  0.104419	$&$  0.103403	$&$  0.002020	$&$  0.002020	$\\
$\delta_{5}$        &$  0.052062	$&$  0.051554	$&$  0.001000	$&$  0.001000	$\\
$\delta_{6}$        &$  0.052285	$&$  0.051776	$&$  0.001001	$&$  0.001000	$\\
$\delta_{7}$        &$  0.010814	$&$  0.009304	$&$  0.000452	$&$  0.000450	$\\
$\delta_{8}$        &$  0.006404	$&$  0.005649	$&$  0.000326	$&$  0.000325	$\\
$\delta_{9}$        &$  0.036142	$&$  0.031612	$&$  0.001797	$&$  0.001788	$\\
$\delta_{10}$       &$  0.006306	$&$  0.005551	$&$  0.000306	$&$  0.000305	$\\
$\delta_{11}$       &$  0.006406	$&$  0.005651	$&$  0.000326	$&$  0.000325	$\\
$\delta_{12}$       &$  0.023824	$&$  0.020805	$&$  0.001144	$&$  0.001139	$\\
$\delta_{13}$       &$  0.004305	$&$  0.003551	$&$  0.000106	$&$  0.000105	$\\
$\delta_{14}$       &$ 10.295675	$&$  7.705024	$&$  0.067351	$&$  0.033683	$\\
$\delta_{15}$       &$  9.463187	$&$  7.086699	$&$  0.043308	$&$  0.021643	$\\
$\delta_{16}$       &$  0.012221	$&$  0.010711	$&$  0.000632	$&$  0.000630	$\\
$\delta_{17}$       &$  0.012321	$&$  0.010811	$&$  0.000652	$&$  0.000650	$\\
$\delta_{\mathtt{I}}$       &$  0.052018	$&$  0.051510	$&$  0.001000	$&$  0.001000	$\\
$\tau_0$            &$  2.469858	$&$  2.465287	$&$  0.009222	$&$  0.009221	$\\
$\alpha$        &$     0.250	$&$     0.031	$&$     0.250	$&$     0.031	$\\
$k_\mathtt{pls}$    &$4		$&$3		$&$6 		$&$3		$\\
$\eta_1$				&$0.031250	$&$0.031250	$&$0.003906	$&$0.031250	$\\
$\varepsilon_{1}$   	&$0.0033	$&$0.0026	$&$6.1e-06	$&$4.7e-06	$\\
$\varrho_{1}$   		&$0.0015	$&$0.0012	$&$0.00074	$&$0.00058	$\\
$\Delta_{C}$   		&$  10	$&$ 7.7	$&$0.067	$&$0.034	$\\
$\Delta_{1}$   		&$     978.4	$&$     732.5	$&$      42.7	$&$       2.7	$\\
\noalign{\smallskip}\hline
\end{tabular}
\end{table}

In the first case (shown as the Case I in the first column of the values in Table~\ref{tab:parameters}, the similar below), the network is configured as $n_0=6$, $f_0=1$, $n_1=3$ and $f_1=1$.
With this, the parameters $\delta_\mathtt{p}=100~\mu s$ and $\delta_\mathtt{d}= 1000~\mu s$ are set with typical message delays that can be supported in common LAN networks shown in Fig.~\ref{fig:network}.
The parameters $\Delta_{0}=1~s$, $\rho=10^{-4}$, $\varepsilon_{0}=1~\mu s$, and $\delta_0=1~s$ can be supported in the most common hardware PTP realizations.
And the parameter $\varepsilon_{2}=50~ms$ can be easily supported with common NTP clients.
But unfortunately, it shows that the expected overall stabilization time $\Delta_{1}$ can be nearly $1000~s$ with these basic system settings.
This is mainly because the basic synchronization cycle is restricted by  $\delta_0$.
So, it is suggested that the updating spans of the underlying CS protocols should be as short as possible.
Another reason for the enlarged stabilization time is that the number of the nodes in $U_0$ is insufficient to minimize $k_\mathtt{pls}$.

In the second case, all system parameters remain the same as the first case except that we set $n_0$ and $f_0$ as $100$ and $3$, respectively.
It is easy to see that the probability of more than $3$ nodes in $100$ independent nodes being simultaneously faulty is very small.
In the table, it shows that with the larger $n_0/f_0$, $\Delta_{1}$ can be reduced with the smaller $k_\mathtt{pls}$.
Also, the synchronization precision and accuracy can be improved with the smaller $\alpha$.
This means that the stabilization can be accelerated, and the synchronization qualities can be improved by deploying more terminal nodes.
However, in comparing the first two cases, these improvements are insignificant.
In these two cases, the final synchronization precision $\varepsilon_{1}$ is coarse if it is compared to the underlying $\mathcal{P}$ protocol.
This is mainly because the synchronization precision is restricted by the indeterminacy (measured as $\delta_\mathtt{d}$ in considering the worst cases) of the processing delays in collecting the remote clock readings.
Another reason is that the clock drifts during a basic synchronization round can be more significant than the errors of remote clock readings.
For example, as the maximal clock drift-rate $\rho$ is set as $10^{-4}$ and the nominal synchronization cycle $\tau_0$ can be in the order of several seconds, the accumulated clock drifts in the convergence process can be at the order of several milliseconds even with the improved convergence rate.

In the third case, the network is configured as $n_0=6$, $f_0=1$, $n_1=5$ and $f_1=2$.
As is discussed in Section \ref{sec:Introduction}, with the larger $f_1$, the reliability and the scalability of the system can be better balanced.
For the system parameters, firstly, with the sub-nanosecond CS protocol WR, we set $\varepsilon_{0}=1~ns$.
As SyncE is also employed in WR, we can accordingly set a much smaller hardware clock drift-rate $\rho=10^{-6}$ (actually can be better than this even without employing SyncE in the typical working environment of PTP \citep{SNLA098A}).
The parameters $\delta_0=50~\mu s$, $\delta_\mathtt{p}=20~\mu s$, and $\delta_\mathtt{d}=100~\mu s$ can be supported in some customized Ethernet \citep{YuCOTS2021}.
And the parameter $\varepsilon_{2}=1~ms$ can be easily supported with some external time resources like GPS clocks.
It shows that the expected overall stabilization time $\Delta_1$ can be greatly reduced with this setting.
Meanwhile, the final synchronization precision and accuracy can also be improved, as they mainly depend on $\rho$, $\delta_\mathtt{d}$, $\alpha$, and $\varepsilon_{0}$.
However, as the synchronization precision provided by the external time reference (like the common NTP clients) is very coarse in comparison to the WR protocol, it needs a significant $k_\mathtt{pls}$ to bring the system from a coarsely synchronized state to the final stabilized precision.
Also, as there are $f_1=2$ faulty networks to be tolerated, the expected stabilization time $\Delta_1$ is enlarged to several ten seconds.

In the last case, we again set $n_0=100$, $f_0=3$, $n_1=3$ and $f_1=1$ as in the second case.
In this case, the synchronization precision $\varepsilon_{1}$ can be improved to the order of several microseconds.
Besides, the expected overall stabilization time $\Delta_{1}$ is reduced to about $3~s$, which can be much faster than the average manual operations.
This is mainly because $f_1$ is reduced to $1$, with which the probability $\eta_1$ can be significantly improved.
Another reason is that $k_\mathtt{pls}$ is minimized to $3$ with the large $n_0/f_0$.

It should be noted that the stabilization time being analyzed here is under the consideration of the worst cases.
In considering many non-worst cases, the average stabilization time can often be much less than $\Delta_{1}$, as is shown in the next section.
Meanwhile, the IS-BFT-CS solution is constructed without utilizing any kind of exact Byzantine agreement.
This can significantly improve the efficiency of the BFT CS system as high message complexity is often required in exact Byzantine agreements.
Compared to other BFT CS protocols that do not rely on the exact Byzantine agreement, the proposed IS-BFT-CS solution can reduce the stabilization time by discreetly utilizing the open-world time resources.
For example, even with $\delta_\mathtt{d}=1~\mu s$, $\rho=10^{-6}$, and omitting all other delays, the expected stabilization time of the original hopping-based SS-BFT-CS \citep{DolevWelchSelf2004} would still be more than five days in tolerating just one Byzantine fault.
With a much-relaxed system setting as the Case IV, the expected stabilization time of the proposed IS-BFT-CS solution is less than three seconds.
This is mainly because the stabilization of the BFT CS system can be significantly accelerated by referencing the temporarily synchronized external clocks when the BFT CS system is not stabilized.

\section{Numerical simulations}
\label{sec:Result}
In the former sections, we have provided a basic IS-BFT-CS solution upon CCBN by integrating the decoupled strong synchronizer, basic detectors, clock merger, and the alien clocks.
Then, this basic solution is analyzed with all worst-case considerations.
Namely, by assuming a malicious adversary who can arbitrarily configure the initial states of the $\mathcal{L}$ system, arbitrarily control the message delays and clock drifts in some bounded ranges, and arbitrarily choose a number of nodes in the $\mathcal{L}$ system being Byzantine, we have shown how the given IS-BFT-CS solution can reach stabilization in considering all worst-case scenarios.
In practice, however, not only the abilities to work under worst-case scenarios but the average performance of the CS systems are of great importance.
Especially in considering average performance in the presence of Byzantine faults, the average stabilization time may also be an essential property.
In this section, we further measure the average stabilization time of the given CS solution with stochastic message delays and uniformly distributed initial systems states.
By doing this, the average property (with stochastic initial system states) can be measured without losing the worst-case consideration for tolerating the Byzantine nodes.

\subsection{Simulation model in measuring average stabilization time}
For simplicity, the $\textit{EoR}$ condition checked in executing the line~\ref{code:corrector_eor} of the $\mathtt{H\_CORRECTOR}$ algorithm can be computed as $\textit{alerted}_j\land \textit{coin}_j$.
With this, the core synchronization process of the IS-BFT-CS solution can be reduced as follows.
Firstly, with the analysis of the strong synchronizer, when there is a desired synchronization point, the stabilization of the $\mathcal{L}$ system would only depend on the tossed coins.
Namely, as there might be some node $j$ still observing $\textit{alerted}_j(t)=1$ when $\mathcal{L}$ is not stabilized, $\textit{coin}_j$ is expect to be $0$ in executing the line~\ref{code:corrector_eor} of the $\mathtt{H\_CORRECTOR}$ algorithm to allow $\mathcal{L}$ to be synchronized with the strong synchronizer.
Secondly, with the analysis of the basic corrector, such kind of synchronization point can also be reached by the tossed coins, as long as no node $j\in U_1$ observes $\textit{alerted}_j(t)=0$.
Thirdly, if some node $j\in U_1$ observes $\textit{alerted}_j(t)=0$, the stabilization of the $\mathcal{L}$ system still only depends on the tossed coins.
Thus, we get that the simulation time can be safely reduced to the discrete instants when some node $j\in U_1$ executes the line~\ref{code:corrector_eor} of the $\mathtt{H\_CORRECTOR}$ algorithm.

With the discrete simulation time, the simulation process can be further separated into three subprocesses.
During the initial subprocess, the $\mathcal{L}$ system is started with an arbitrary initial state (being simulated with uniformly distributed local clocks for our aim here).
Then, the randomized initial subprocess proceeds until some desired system states appear with the desired synchronization point and the current coins being tossed in the desired way, with which the deterministic convergence subprocess starts.
Then, when all nodes in $U_1$ observe $\textit{alerted}_j(t)=0$, the simulation process enters the deterministic stabilized subprocess.
Thus, the measurement performed here is to count the time passed in the first two subprocesses during every simulation process.

\subsection{Simulation results}

\begin{figure}[htbp]
\centering
\begin{subfigure}{.23\textwidth}
\centering\includegraphics[width=1.7in]{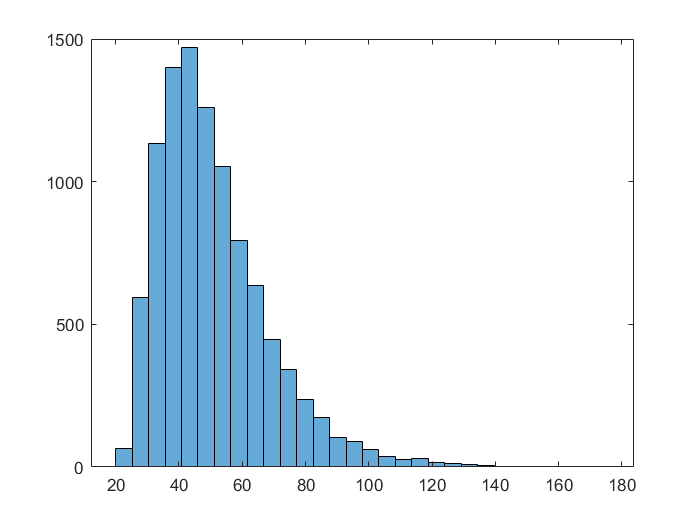}
\caption{The distribution}
\label{fig:simu1}
\end{subfigure}
\begin{subfigure}{.23\textwidth}
\centering\includegraphics[width=1.7in]{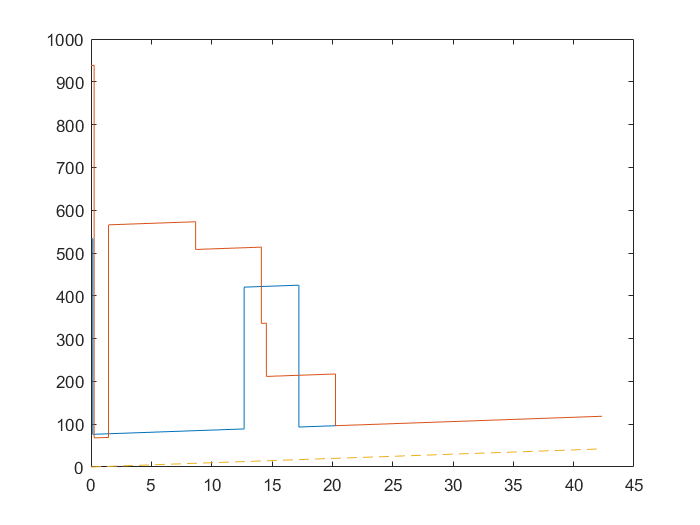}
\caption{An instance}
\label{fig:simu2}
\end{subfigure}
\caption{Average stabilization time with system setting I}
\label{fig:simu12}
\end{figure}

\begin{figure}[htbp]
\centering
\begin{subfigure}{.23\textwidth}
\centering\includegraphics[width=1.7in]{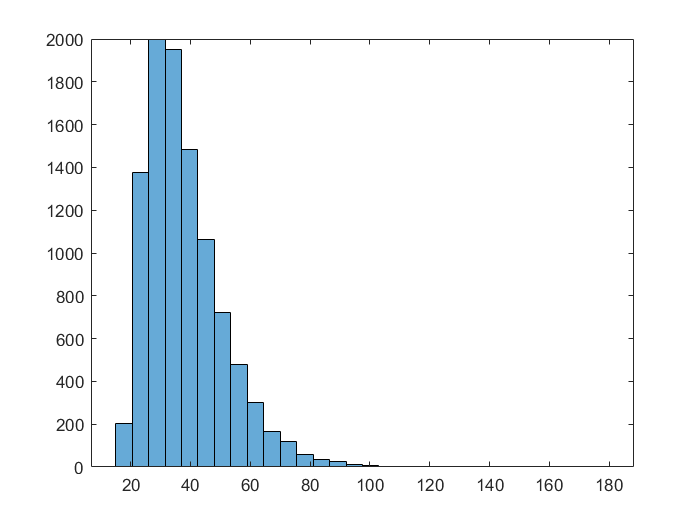}
\caption{The distribution}
\label{fig:simu3}
\end{subfigure}
\begin{subfigure}{.23\textwidth}
\centering\includegraphics[width=1.7in]{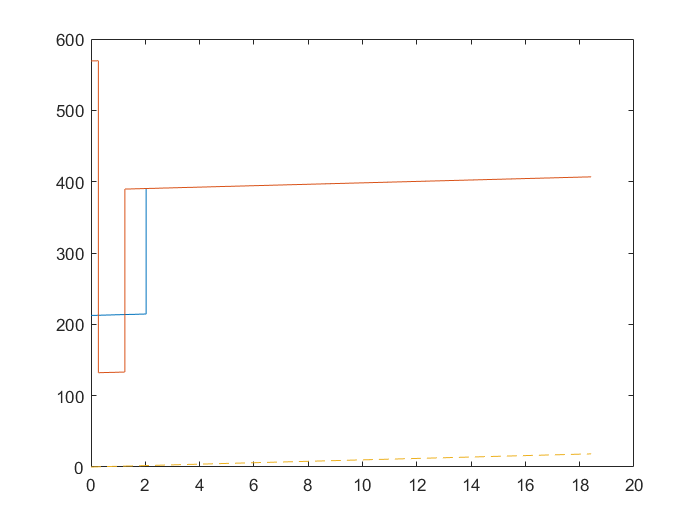}
\caption{An instance}
\label{fig:simu4}
\end{subfigure}
\caption{Average stabilization time with system setting II}
\label{fig:simu34}
\end{figure}

\begin{figure}[htbp]
\centering
\begin{subfigure}{.23\textwidth}
\centering\includegraphics[width=1.7in]{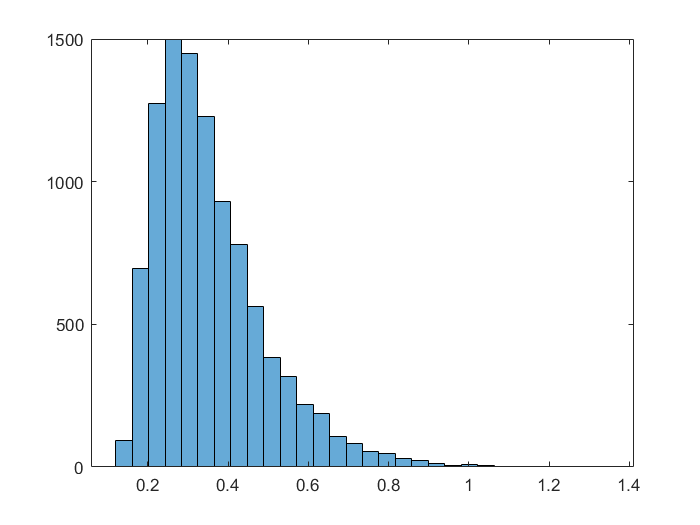}
\caption{The distribution}
\label{fig:simu5}
\end{subfigure}
\begin{subfigure}{.23\textwidth}
\centering\includegraphics[width=1.7in]{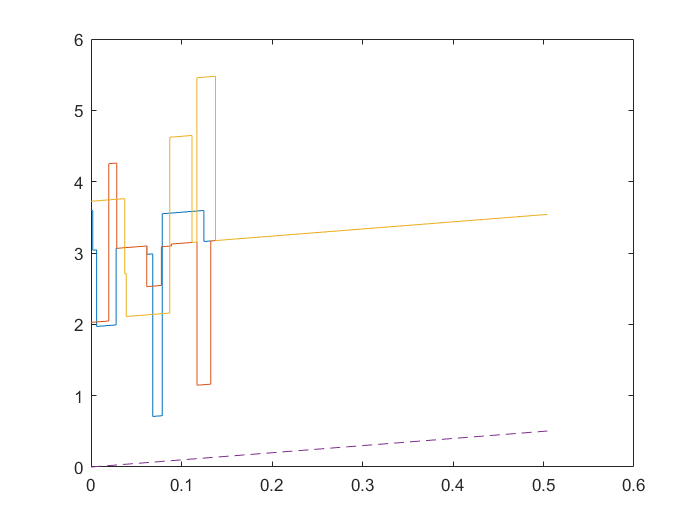}
\caption{An instance}
\label{fig:simu6}
\end{subfigure}
\caption{Average stabilization time with system setting III}
\label{fig:simu56}
\end{figure}

\begin{figure}[htbp]
\centering
\begin{subfigure}{.23\textwidth}
\centering\includegraphics[width=1.7in]{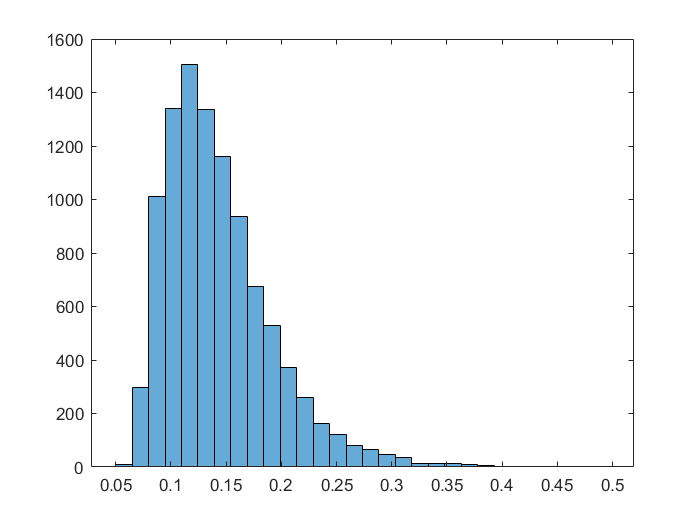}
\caption{The distribution}
\label{fig:simu7}
\end{subfigure}
\begin{subfigure}{.23\textwidth}
\centering\includegraphics[width=1.7in]{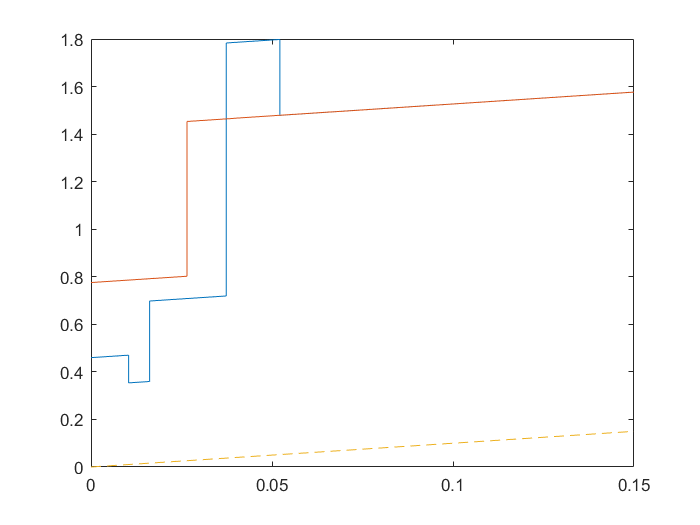}
\caption{An instance}
\label{fig:simu8}
\end{subfigure}
\caption{Average stabilization time with system setting IV}
\label{fig:simu78}
\end{figure}

The simulation results of the four system settings corresponding to the four cases of Table~\ref{tab:parameters} are shown in Fig.~\ref{fig:simu12} to Fig.~\ref{fig:simu78}, respectively.
For every system setting, the collected distribution (of $10000$ instances) of the stabilization time (still being measured in seconds) and a randomly chosen simulation process are respectively shown in the left and right subfigures.

In Fig.~\ref{fig:simu12}, the stabilization time is simulated with the system setting I (Case I of Table~\ref{tab:parameters}, the similar below).
It shows that although the expected stabilization time is about $1000$ seconds in considering the worst-case initial system state, the average result can be much better.
This is mainly because that some very special cases in the worst-case consideration are very unlikely encountered in some real-world stochastic environment.

In Fig.~\ref{fig:simu34}, the stabilization time is simulated with the system setting II.
It is easy to see that the average stabilization time can also be reduced by deploying more terminal nodes in improving $\alpha$.

In Fig.~\ref{fig:simu56}, the stabilization time is simulated with the system setting III.
In comparing the average stabilization time with existing solutions, the state-of-the-art randomized SS-BFT-CS solution FATAL proposed in \cite{Dolev2014Rigorously,Dolev2014PulseGeneration} achieves the average stabilization time of several seconds (about $5~s$) in the presence of two Byzantine nodes in CCN without employing exact Byzantine agreement and external time resources.
Here, by setting $f_1=2$ in Case III, the average stabilization time reached in the provided IS-BFT-CS solution (less than $1~s$) is much shorter than FATAL \citep{Dolev2014Rigorously}.
It should also be noted that the experimental results reported in \cite{Dolev2014Rigorously} are given in the background of tiny-sized Systems-on-Chips (SoCs), with which the basic synchronization cycles are often less than one microsecond.
In our cases, the basic synchronization cycles are much larger.
So, it is shown that by discreetly utilizing the available external time in IoT systems, the average stabilization time can be greatly reduced in comparing with traditional randomized SS-BFT-CS solutions without employing exact Byzantine agreement.
It should be noted that, as the Byzantine faults are hard to be well-generated in experimental environments, the given results integrates the analysis of the deterministic aspect of the BFT algorithms and the simulation of the stochastic aspect of the randomized algorithm.
Comparing with the experimental results \citep{Dolev2014Rigorously}, the Byzantine faults are more safely handled with the reduced simulation model.

Lastly, in Fig.~\ref{fig:simu78}, the stabilization time is simulated with the system setting IV.
Comparing Case IV with Case III, it shows that although the worst-case stabilization time can be greatly reduced with a smaller $f_1$, the average performance of the system with a slightly larger $f_1$ is not much worse than the case $f_1=1$.
This is mainly because some extreme conditions that may exponentially increase the stabilization time are very unlikely satisfied with stochastic initial system states.
However, in considering the worst-case scenarios, these extreme conditions can be satisfied, and thus the expected stabilization time would be significantly enlarged.
This is the main difference between the average properties and the worst-case ones considered in the former section.

\section{Conclusion}
\label{sec:Conclusion}

In this paper, we have investigated the IS-BFT-CS problem and provided an IS-BFT-CS solution upon heterogeneous IoT networks.
Firstly, by abstracting the LAN-layer networks as CCBN and providing the minimized safe interface for the two sides, the IS-BFT-CS problem is identified in the context of the open-world networks.
With this, the basic IS-BFT-CS solution is provided upon CCBN, which utilizes the open-world time resources as temporary synchronized external clocks (i.e., the alien clocks) for achieving faster stabilization.
Meanwhile, for better integrating the distributed BFT-CS and the master-slave CS, we have presented a modularized framework and provided the IS-BFT-CS solution with decoupled building blocks.
In measuring the properties of the provided solution, formal analysis and numerical simulations are successively presented.

In the practical perspective, we have shown that with several arbitrarily connected heterogeneous (or homogeneous) communication subnetworks, some reliable, efficient, and high-precision ICS systems can be built upon CCBN by integrating the common high-precision server-client CS and the traditional ultra-high reliable distributed CS with discreet use of the external time references.
Also, in considering the various real-world and future IoT applications, we have shown that different kinds of underlying CS protocols can be utilized under the same IS-BFT-CS framework with reusable building blocks (such as the synchronizers, the detectors, the clock mergers).
In the theoretical perspective, we have shown that intro-stabilization provides a discreet way to integrate traditional BFT algorithms with some new open-world resources.
Meanwhile, only $n_1>2f_1$ is required in the provided IS-BFT-CS solution upon CCBN, which outperforms the traditional Byzantine resilience for reaching self-stabilization in CCN.

Despite the merits, the provided IS-BFT-CS solutions can be further improved in several ways.
Firstly, the CCBN network model might over-abstract real-world large-scale IoT systems.
Future IS-BFT-CS solutions can be developed upon multi-layer CCBN and even sparsely connected bipartite networks for better scalability.
For example, we can build a multi-layer IS-BFT-CS system where the manager nodes in each such intro-stabilizing layer would establish their alien clocks by referencing to the clocks of upper layer nodes.
Also, in constructing the IS-BFT-CS solutions with the external time, the algorithms provided in this paper are rather heuristic than optimal in stabilization time, message complexity, and synchronization precision.
Moreover, in providing external time service, the IS-BFT-CS solutions should be further safely integrated with ECS solutions.

\bibliographystyle{IEEEtran}
\bibliography{IEEEabrv,IOT}

\end{document}